\documentclass[twocolumn]{article}

\usepackage{amssymb}
\usepackage{amsmath,array} 
\usepackage{graphicx} 
\usepackage{subfig}
\usepackage{hyperref}

\usepackage{color}
\newcommand{\RR}[1]{\mathbb{R}^{#1}} 
\newcommand{\bd}[1]{\mbox{\boldmath $#1$}}

\usepackage{authblk}

\usepackage[left=1.5cm,right=1.5cm,top=2.0cm,bottom=2.5cm,includeheadfoot]{geometry}

\title{\vspace{-2.0\baselineskip}Model-Coupled Autoencoder for Time Series Visualisation}
\author[1]{Nikolaos Gianniotis}
\author[1]{Sven D. K\"ugler}
\author[1]{Peter Ti\v no}
\author[2]{Kai L. Polsterer}
\affil[1]{Astroinformatics Group, Heidelberg Institute for Theoretical Studies (HITS), Schloss-Wolfsbrunnenweg 35, 69118 Heidelberg, Germany}
\affil[2]{School of Computer Science, The University of Birmingham, Edgbaston, Birmingham B15 2TT, United Kingdom}

\date{}
\begin{document}

\maketitle

\begin{abstract}
We present an approach for the visualisation of a set of time series that combines an echo state network with an autoencoder. For each time series in the dataset we train an echo state network, using a common and fixed reservoir of hidden neurons, and use the optimised readout weights as the new representation. Dimensionality reduction is then performed via an autoencoder on the readout weight representations. The crux of the work is to equip the autoencoder with a loss function that correctly interprets the reconstructed readout weights by associating them with a reconstruction error measured in the data space of sequences. This essentially amounts  to measuring the predictive performance that the reconstructed readout weights exhibit on their corresponding sequences when plugged back into the echo state network with the same fixed reservoir.
We demonstrate that the proposed visualisation framework can deal both with real valued sequences as well as
binary sequences. We derive magnification factors in order to analyse distance preservations and distortions in the visualisation space.
The versatility and advantages of the proposed method are demonstrated on datasets of time series that originate from  diverse domains.
\end{abstract}

\section{Introduction}
\label{sec:introduction}

Time series\footnote{We  interchangeably use the terms time series and sequence.} are sequences of observations that exhibit  short or long term
 dependencies between them in time. These dependencies can be thought of as  manifestations of a latent regime (e.g.~natural law) governing the behaviour of the time series. Machine learning approaches  designed to deal with data of a vectorial nature have no knowledge of such temporal dependencies. By introducing a model that accounts for  temporal behaviour,  algorithms  can become ``aware'' of the sequential nature of the data and make the most of the available information.

Echo state networks (ESNs) \cite{Lukosevicius2009} are discrete time recurrent neural networks that have emerged as a popular method to capture the latent regime underlying a time series.
ESNs have the great advantage that the hidden part, the reservoir of neurons,
is fixed and only the output weights need to be trained. The ESN is thus essentially a linear model and so the output weights, also known as readout weights, can thus be easily optimised via least squares.
The processing of structured data has been a topic of research for a long time \cite{Jaakkola1998, Jebara2004}. Regarding time series, recent attempts \cite{Chatzis2011, Chen2013, Chen2014} have exploited the predictive capabilities of ESNs in regression and classification tasks. 
In the unsupervised setting, the work in \cite{Sperduti2013} suggests compressing a  linear state space model through a linear autoencoder in order to extract vectorial representations of structured data. The work in \cite{Wang2012} considers the visualisation of  individual observations belonging to a {\itshape single} sequence by temporally linking them using an ESN.

In this work, we  employ the ESN in the formulation of a dimensionality reduction algorithm for visualising a dataset of time series (we extend previous work presented in \cite{Gianniotis2015}).
Given a fixed reservoir, the only free parameters in the ESN are in the readout weight vector which maps the state space to the sequence space.
Thus, an optimised (i.e.~trained) readout weight vector  uniquely addresses an instance of the ESN (always for the same fixed reservoir) that best predicts on  a given sequence. 
We can also reason backwards: given an observed sequence, we can train the ESN (Section \ref{sec:echo_state_networks}) and identify the readout weight vector that best predicts the given sequence.
Hence, each sequence in the dataset can be mapped to the readout weight vector that exhibits the best predictive performance on it.  These readout weight vectors in conjunction with the common and fixed reservoir, capture temporal features of their respective sequences. Representing sequences as weight vectors, constitutes the first part of our proposed approach (Section \ref{sec:encoding}).

\begin{figure*}
\centering
\includegraphics[trim=0 0 0 0.2cm, width=0.90\textwidth]{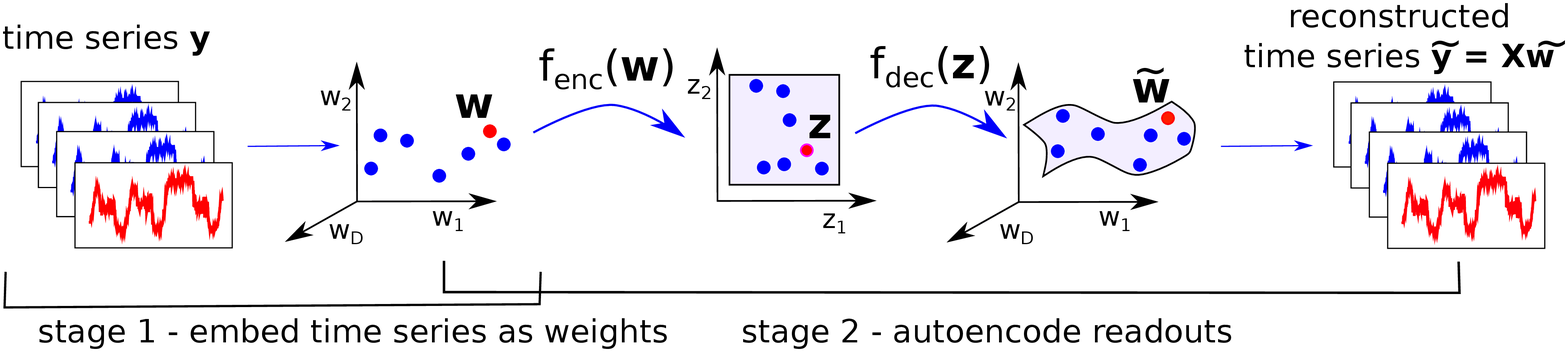}
\caption{Sketch of proposed method. In a first stage, time series \bd{y} are cast to readout weights  $\bd{w}$ in the weight space (see Section \ref{sec:encoding}). In a second stage, the autoencoder projects readout weights  $\bd{w}$ onto coordinates $\bd{z}$ residing in a two-dimensional space,
and reconstructs them again as $\tilde{\bd{w}}$ (see Section \ref{sec:esn_coupled}).
By multiplying with the state space, given by \bd{X}, we map the reconstructed readout weights $\tilde{\bd{w}}$ to the sequence space
where reconstruction error is measured (see Eq.~\eqref{eq:new_objective}).
}
\label{fig:overview}
\end{figure*}

The second stage of our approach involves training an autoencoder  \cite{Kramer1991} on the obtained readout weight vectors in order to induce a two-dimensional representation, the visualisation, at the bottleneck. At the heart of the autoencoder lies the  reconstruction error function  
which drives the visualisation induced at the bottleneck.
During training, the autoencoder receives readout weights as inputs, compresses them at the bottleneck, and outputs an approximate version of the inputs, the reconstructed readout weights.
Typically, one would take as the reconstruction error function the $L_2$ norm between the original readout weights and reconstructed readout weights. In the proposed work, we equip the autoencoder with a different reconstruction function
that assesses how well the reconstructed readout weights still predict on the sequence that it represents.
If it predicts well, we deem it a good reconstruction; if it predicts poorly, we deem it a poor reconstruction
(Section \ref{sec:esn_coupled}). An overview of the proposed method is displayed in Fig.~\ref{fig:overview}.

In Section \ref{sec:experiments}, we show that the autoencoder with the proposed reconstruction error function is capable of interpreting similarities between time series better than other dimensionality reduction algorithms. In Section \ref{sec:discussion}, we discuss 
the possibility of alternative formulations of the proposed approach
before concluding with some remarks on future work in Section \ref{sec:conclusion}.

\section{Preliminary}
\label{sec:Preliminary}

This section introduces some notation and terminology while briefly reviewing ESNs and the autoencoder.

\subsection{Echo State Networks}
\label{sec:echo_state_networks}

An ESN is a discrete-time recurrent neural network with fading memory. It processes time series
composed by a sequence of observations $y(t)\in \RR{}$  over time $t$ that we  denote here by $\bd{y}=(y(1),\dots,y(T))$, where $T$ is the length\footnote{In general, each sequence can have its own length $T_n$. For ease of exposition, here all sequences have the same $T$.} of the sequences. Hence $\bd{y} \in \RR{T \times 1}$.
Given an input $y(t)$, the task of the ESN is to make a prediction $\hat{y}(t+1)$ for the observation $y(t+1)$ in the next time step.
Similarly to a feedforward neural network, the ESN comprises an input layer with weights $\bd{v}\in\RR{D\times 1}$, a hidden layer with weights $\bd{U}\in\RR{D\times D}$ (hence $D$ is the size of the reservoir) and an output layer with weights $\bd{w}\in\RR{D\times 1}$, the latter weights $\bd{w}$ also known as readout weights. 
However, in contrast to feedforward networks, ESNs equip the hidden neurons with feedback connections. The operation of an ESN is specified by the equations:
\begin{align}
\bd{x}(t+1) &=h(\bd{U}\bd{x}(t) + \bd{v} y(t)) \ , \label{eq:esn_recursive_one} \\
\hat{y}(t+1) &=  \bd{w}^T \bd{x}(t+1) \label{eq:esn_recursive_two} \ ,
\end{align}
where $\bd{x}(t)\in\RR{D\times 1}$ are  the hidden activations of the reservoir at time $t$, and $h(\cdot)$ is a nonlinear function commonly chosen as the $\tanh(\cdot)$ function. Bias terms have been omitted in the formulation for the sake of clarity in notation.

According to standard ESN methodology \cite{Lukosevicius2009}, parameters  \bd{v} and \bd{U} in Eqs.~\eqref{eq:esn_recursive_one}, \eqref{eq:esn_recursive_two} are randomly
generated\footnote{The spectral radius of the reservoir's
weight matrix \bd{U} is made $< 1$ to encourage {\it the Echo State Property}.} and fixed. The only trainable parameters are the readout weights \bd{w}. 
Training involves feeding at each time step $t$ an observation $y(t)$ and recording
the resulting activations $\bd{x}(t)$ row-wise into a matrix $\bd{X}\in\RR{T \times D}$. Usually, some initial
observations are dismissed in order to ``washout" \cite{Lukosevicius2009} dependence on the initial arbitrary reservoir state (e.g. $\bd{x}(1)=\bd{0}$). Given matrix $\bd{X}$, the following objective function is minimised:
\begin{equation}
\ell(\bd{w}) =\|\bd{X}\bd{w} - \bd{y}\|^2 \ .
\label{eq:ESN_objective}
\end{equation}
The above objective
can be supplemented by a regularisation term and so the combined objective is $\ell(\bd{w}) + \mu^2 \|\bd{w}\|^2 $. 
The combined objective can be exactly minimised by solving the pertaining least squares problem and obtaining
$\bd{w} = (\bd{X}^T \bd{X}+ \mu^2 \bd{I}_D)^{-1} \bd{X}^T  \bd{y} $ as the solution,
where $\bd{I}_D$ is the $D\times D$ identity matrix. 
Given this result, we introduce function $g(\bd{y})$ that maps a given time series to the optimal readout weights:
\begin{equation}
g(\bd{y}) = (\bd{X}^T \bd{X} + \mu^2 \bd{I}_D)^{-1} \bd{X}^T  \bd{y} = \bd{w} \ .
\label{eq:esn_project}
\end{equation}

\subsection{Deterministically Constructed Echo State Networks}
\label{sec:deterministic}

In the original formulation of the ESN \cite{Lukosevicius2009} the weights in \bd{v} and \bd{U} are generated stochastically and so are the connections between the hidden neurons in the reservoir. This makes the training and use of the ESN dependent on random initialisations. In order to avoid this source of randomness, we make use of a class of ESNs that are constructed in a deterministic fashion \cite{Rodan2011}.

\begin{figure}
\centering
\includegraphics[width=0.22\textwidth]{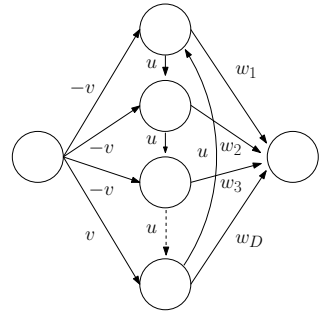}
\caption{Deterministic ESN with cyclic architecture, see Section \ref{sec:deterministic}. Circles denote  neurons and arrows connections between neurons.
All input neurons connect to the hidden neurons, and all hidden neurons connect to the output neurons. Hidden neurons are connected in a cyclic fashion to each other. All input weights have the same absolute value $v$, and the sign is determined by a deterministic aperiodic sequence. The hidden reservoir weights are fixed to the same scalar $u$. The readout weights $\bd{w}$ are the only adaptable part of the ESN. }
\label{fig:cyclic_esn}
\end{figure}

Deterministic ESNs make several simplifications over standard ESNs. All entries in  $\bd{v}$ have the same absolute value of a single scalar parameter $v>0$.
The signs of the entries in $\bd{v}$ are deterministically generated by an aperiodic sequence: e.g. a pseudorandom binary sequence (coin flips), with  outcomes $0$ and $1$ corresponding to $-$ and $+$ respectively.
Similarly, the entries in $\bd{U}$ are  parametrised by a single scalar $u>0$. As opposed to random connectivity, deterministic ESNs impose a fixed regular topology on the hidden neurons in the reservoir. Amongst possible choices, one can arrange the neurons in a cycle. A cyclic arrangement imposes the following structure on  $\bd{U}$: the only nonzero entries in $\bd{U}$ are on the lower sub-diagonal
$\bd{U}_{i+1,i} = u$, and at the upper-right corner $\bd{U}_{1,D} = u$.
An illustration of a cyclic deterministic ESN is shown in Fig.~\ref{fig:cyclic_esn}.

In this work we employ deterministic ESNs with a cyclic connectivity. Deterministic ESNs have three degrees of freedom: the reservoir size $D$, the input weight $v$ and  reservoir weight $u$. Hence, the triple $(D,v,u)$ completely specifies an ESN.  
It has been shown empirically and theoretically (memory capacity) \cite{Rodan2011}   that deterministic ESNs perform up to par with their stochastic counterparts.
Training of a deterministic ESN is performed in exactly  the same fashion as in  stochastically constructed ESNs  using the objective $\ell(\bd{w})$ in Eq.~\eqref{eq:ESN_objective}.

\subsection{Autoencoder}
\label{sec:autoencoder}

The autoencoder \cite{Kramer1991} is a feedforward neural network that defines
 a three hidden layer architecture\footnote{To be perfectly precise, we use what is widely considered the standard autoencoder specified  in  \cite[Sec. 12.4.2]{Bishop}).} with the middle layer, the  ``bottleneck'', being smaller than the others in terms of the number of neurons denoted by $Q$. 
The autoencoder learns  an identity mapping by training on targets identical to the inputs.
Learning is hampered by the bottleneck that forces the autoencoder
to reduce the dimensionality of the inputs, and hence the output is only an
approximate reconstruction of the input.

Given general vectors $\bd{s}$, we want to reduce them to a $Q$-dimensional representation. 
The autoencoder is the composition of an encoding $f_{enc}$ and a decoding $f_{dec}$
function. Encoding maps inputs $\bd{s}$ to low-dimensional compressed versions, $f_{enc}(\bd{s}) = \bd{z} \in\RR{Q}$, while decoding  maps approximately back to the inputs, $f_{dec}(\bd{z}) = \tilde{\bd{s}}$. The complete autoencoder is the function $f(\bd{s};\bd{\theta}) = f_{dec}(f_{enc}(\bd{s})) = \tilde{\bd{s}}$, where \bd{\theta} are the weights of the autoencoder. Training the autoencoder involves minimising the $L_2$ norm between  $N$ given vectors $\bd{s}$  and their reconstructions:
\begin{equation}
\sum_{n=1}^{N} \| \tilde{\bd{s}}_n - \bd{s}_n\|^2  = \sum_{n=1}^{N} \| f(\bd{s}_n;\bd{\theta}) - \bd{s}_n\|^2  .
\label{eq:L2_AE_objective}
\end{equation}
%

\section{Model Formulation}
\label{model_formulation}

The proposed approach consists of two stages. In Section \ref{sec:encoding}, we
discuss how time series $\bd{y}$ are embedded in the space of readout weight vectors $\bd{w}$.
Section \ref{sec:esn_coupled} discusses how an autoencoder with a modified reconstruction function
is applied on the readout weight vectors in a meaningful manner.

\subsection{Embedding time series in the space of readout weights}
\label{sec:encoding}

Given a deterministically constructed and fixed reservoir $(D,v,u)$, we cast a  dataset $\{\bd{y}_1,\dots,\bd{y}_N\}$ 
via $g(\bd{y}_n) = \bd{w}_n$ to a  new dataset of readout weights $\{\bd{w}_1,\dots,\bd{w}_N\}$.
\emph{We emphasise that all time series are embedded in the space of readout weights with respect to the same fixed dynamic
reservoir $(D,v,u)$}. After this embedding,  visualisation proceeds by performing dimensionality reduction on the new representations $\bd{w}_n$.
We take the view that the readout  weight $\bd{w}_n$ is a good representation for a sequence $\bd{y}_n$ with respect to the fixed reservoir $(D,v,u)$. The readout weight $\bd{w}_n$ captures important information about $\bd{y}_n$ in the sense that it exhibits good predictive power on it. Moreover, the readout weight vector $\bd{w}_n$ features  time-shift invariance, and can  accommodate sequences of variable length.

A prerequisite for a successful embedding is a common, fixed reservoir that  enables good predictive  performance on the data. To find this reservoir, we opt for a simple strategy. 
For both $v$ and $u$ we take a regular grid of e.g.~$10$ candidate values $[10^{-2},\dots,1.0]$. 
For each combination  $(u,v) \in [10^{-2},\dots,1.0] \times [10^{-2},\dots,1.0]$, 
we perform the following:
\begin{enumerate}
\item Split each sequence \bd{y} in two halves  $\bd{y}_n^{(train)}$ and  $\bd{y}_n^{(test)}$.
\item According to Eq.~\eqref{eq:ESN_objective}, train  ESN on $\bd{y}_n^{(train)}$ by minimising $\ell^{(train)}(\bd{w}) =\|\bd{X}_n^{(train)}\bd{w} - \bd{y}_n^{(train)}\|^2$ which yields $\bd{w}_n$.
\item Measure test error  via $\ell^{(test)}(\bd{w}_n) =\|\bd{X}_n^{(test)}\bd{w}_n - \bd{y}_n^{(test)}\|^2$.
\end{enumerate}
Matrices $\bd{X}_n^{(train)}$ and $\bd{X}_n^{(test)}$ respectively record row-wise the activations $y_n^{(train)}(t)$ and $y_n^{(test)}(t)$  as specified in Section \ref{sec:echo_state_networks}.
The combination $(u,v)$ with the lowest test error over all sequences $\sum_{n=1}^N \ell^{(test)}(\bd{w}_n)$, determines the ESN that will cast all time series in the dataset to readout weights.
Parameters $D$ and $\mu$ may also be included in this simple validation  scheme.

\subsection{ESN-coupled Autoencoder} 
\label{sec:esn_coupled}

We want to reduce the dimensionality of the new representations  $\{\bd{w}_1,\dots,\bd{w}_N\}$
 using an autoencoder. One possibility is to  directly apply the autoencoder taking as input the readout weights and returning 
their reconstructed versions, $f: \RR{D\times 1} \rightarrow \RR{D\times 1}$.
We could then minimise the following objective function with respect to the autoencoder weights $\bd{\theta}$:
\begin{equation}
 \sum_{n=1}^{N} \| f(\bd{w}_n;\bd{\theta}) - \bd{w}_n\|^2 =  \sum_{n=1}^{N} \| \tilde{\bd{w}}_n - \bd{w}_n\|^2 .
\label{eq:direct_L2_objective}
\end{equation}
A limitation of the above objective function is that it merely measures how well the reconstructions $f(\bd{w};\bd{\theta}) = \tilde{\bd{w}}$ approximate the original inputs $\bd{w}$ in the $L_2$ sense. 

A better objective would measure reconstruction error in the sequence space as opposed to the space of readout weights.
To that purpose, we  map the reconstructed readout weights $\tilde{\bd{w}}$ to the sequence space by multiplying
with the respective state matrix,  $\bd{X}\tilde{\bd{w}} = \tilde{\bd{y}}$.
In actual fact,  function $\ell(\cdot)$ in Eq.~\eqref{eq:ESN_objective} is cut out for this task:
if $\ell(\tilde{\bd{w}})$ returns high likelihood, then $\tilde{\bd{w}}$ is a good reconstruction; if $\ell(\tilde{\bd{w}})$ returns low likelihood, then $\tilde{\bd{w}}$ is a poor reconstruction. The new objective function reads:
\begin{equation}
\sum_{n=1}^{N} \ell_n( f(\bd{w}_n;\bd{\theta})) = 
\sum_{n=1}^{N} \|\bd{X}_n f(\bd{w}_n;\bd{\theta}) - \bd{y}_n\|^2 = 
\sum_{n=1}^{N} \|\tilde{\bd{y}}_n - \bd{y}_n\|^2 \ ,  
\label{eq:new_objective}
\end{equation}
where $\ell_n$  and $\bd{X}_n$ are respectively the objective function and state space pertaining to sequence $\bd{y}_n$, see Eq.~\eqref{eq:ESN_objective}.
The gradient of the new objective function in Eq. \eqref{eq:new_objective} with respect to the weights \bd{\theta}, is calculated by backpropagation \cite{Bishop}. We use L-BFGS as the optimisation routine
for training the weights \bd{\theta}. 


%

\subsection{Data Projection}

Having trained the autoencoder $f(\bd{w}_n;\bd{\theta})$, we would like to project a time series $\bd{y}^*$ to a coordinate $\bd{z}^* \in \RR{Q}$.
To that end, we first use the fixed ESN reservoir to cast the time series to  $g(\bd{y}^*)=\bd{w}^*$.
Then, the readout weight $\bd{w}^*$ is projected using the encoding part of the autoencoder
to obtain  $f_{enc}(\bd{w}^*)= \bd{z}^*$.

\section{Binary Sequences}
\label{sec:symbolic}

The time series considered so far are sequences of reals $y(t)\in \RR{}$. However, it is possible to extend the proposed approach to the processing of symbolic sequences. In particular, we consider binary sequences composed of observations $y(t)\in \{ 0,1\}$. 
For an ESN to process binary sequences, we pass its outputs  through the logistic function $\sigma(\cdot) = (1+\exp(\cdot))^{-1}$  (link function of the Bernoulli distribution).
Hence, the  equations specifying the operation of the ESN now read\footnote{ In Eq.~\eqref{eq:logistic_esn_recursive_one} we subtract 0.5 from $y(t)$, since the symmetric $\tanh(\cdot)$ transfer function $h$ is used in the dynamic reservoir. }:
\begin{align}
\bd{x}(t+1) &=h(\bd{U}\bd{x}(t) + \bd{v}(y(t)-0.5)) \ , \label{eq:logistic_esn_recursive_one} \\
\hat{y}(t+1) &=\sigma(  \bd{w}^T  \bd{x}(t+1)  )\label{eq:logistic_esn_recursive_two} \ .
\end{align}

Here the output $\hat{y}(t+1)\in[0\dots 1]$ of the ESN is interpreted as the probability of the next observation
$y(t+1)$ being equal to $1$, i.e. $\hat{y}(t+1)=p(y(t+1)=1)$.
Accordingly, the objective function $\ell(\bd{w})$ in Eq.~\eqref{eq:ESN_objective}
needs to be redefined. Instead of solving a least squares problem, we minimise the cross-entropy:
\begin{equation}
\ell^{ce}(\bd{w}) =  - \sum_{t=1}^T   y(t) \log \hat{y}(t) \ .
\label{eq:logistic_ESN_objective}
\end{equation}
%
%
Training of the ESN is carried out by iterative gradient minimisation of Eq.~\eqref{eq:logistic_ESN_objective}
preceded by a period of washout.

The above modifications  to the ESN, call for a modification also in the autoencoder. While in Eq.~\eqref{eq:new_objective}
reconstruction  is measured via the least-squares based function $\ell(\bd{w})$, we now use
the cross-entropy based function $\ell^{ce}(\bd{w})$.  In order for the autoencoder to  process correctly the weights coming from binary sequences,
its objective function needs to be changed to:
\begin{equation}
\sum_{n=1}^{N} \ell_n^{ce} ( f(\bd{w}_n;\bd{\theta})) = 
- \sum_{n=1}^{N} \sum_{t=1}^T   y(t) \log \sigma(  f(\bd{w}_n;\bd{\theta})^T \bd{x}_n(t+1)  ) \  .
\label{eq:logistic_new_objective}
\end{equation}
In the case of binary sequences, the outputs of the autoencoder $f(\bd{w};\bd{\theta})$ are put though the function $\ell^{ce}(\cdot)$. 

By adopting a 1-of-K coding scheme for the symbols, and the softmax function in the place of the logistic function, an extension to $K$ number of symbols is possible.  The resulting objective for training the ESN is again a cross-entropy function.

\section{Magnification Factors}
\label{sec:magnification}

\begin{figure}
\centering
\includegraphics[width=0.45\textwidth]{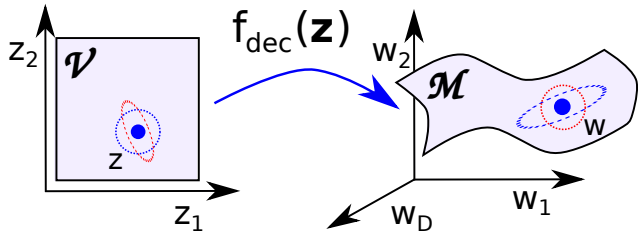}
\caption{Stylised sketch: mapping $f_{dec}(\bd{z})$ embeds the visualisation space $\mathcal{V}$  as a manifold $\mathcal{M}$
in the  space of readout weights.
Each point $\bd{z}$ addresses a probabilistic ESN
with readout weights $\bd{w}$.}
\label{fig:magn_factors}
\end{figure}

In Fig.~\ref{fig:magn_factors}, the  smooth nonlinear function $f_{dec}(\bd{z})$ embeds the low-dimensional visualisation space ${\cal V}$
as a  $Q$-dimensional manifold $\mathcal{M}$ in the space of readout weights $\bd{w}$.
Each point $\bd{z}\in{\cal V}$ addresses an ESN model\footnote{We always have the same fixed reservoir.} with readout weights $\bd{w}\in \mathcal{M}$.
The ESN model may be viewed as a probabilistic model, if we assume that observations $y(t)$ are corrupted by  i.i.d. Gaussian noise of zero mean and variance $\epsilon^2$:
\begin{align}
 \bd{y} =& \bd{X}\bd{w} + \epsilon \ , \\
 p(\bd{y};\bd{w}) =& \mathcal{N}(\bd{y}| \bd{X}\bd{w} ,\epsilon\bd{I}_{T}) \ ,
\end{align}
Thus,  each  point $\bd{z}$ addresses a probabilistic  model $p(\bd{y} ; f_{dec}(\bd{z}))$, and  $\cal M$ is  a  manifold of probabilistic models $p(\bd{y} ; f_{dec}(\bd{z})) $.

Manifold $\cal M$ is endowed with a natural metric for measuring distances between probabilistic models $p(\bd{y} ; f_{dec}(\bd{z})) $. Specifically, the metric tensor on a statistical manifold at a given point $\bd{z}$ is the $Q \times Q$ Fisher information  matrix (FIM)  \cite{Kullback1959}. Here, we approximate it through the {\it observed FIM} over the given dataset of sequences:
\begin{equation}
\bd{F}(\bd{z})_{i,j}\!= \! -\sum_{n=1}^N \left[ \left(\frac{\partial}{\partial z_i} \log p(\bd{y}_n ; f_{dec}(\bd{z}))\right) \left(\frac{\partial}{\partial z_j} \log p(\bd{y}_n ; f_{dec}(\bd{z}))\right) \right] \\ .
\end{equation}

We note that the visualisation space $\cal V$ does not necessarily reflect  distances between models on $\cal M$.
In Fig.~\ref{fig:magn_factors}, we see how neighbourhoods of \bd{z}, depicted as dotted ellipses, transform on $\cal M$.  
Thus, in order to interpret  distances  in $\cal V$, it is important to push-forward the natural notion of distances on $\cal M$ onto the visualization space $\cal V$. In the topographic mapping literature the induced metric in the visualization space from the data space is usually represented through magnification factors \cite{Bishop1997}. In the following we show how magnification factors can be computed in the ESN-AE setting.

Given the FIM, one can push forward local distances $\Delta\bd{z}$ from $\cal M$ onto $\cal V$  via $\Delta\bd{z}^T \bd{F}(\bd{z}) \Delta\bd{z}$. In particular, at a given point $\Delta\bd{z}$ it is possible to estimate in which direction $d\bd{z}$ the distance changes the most. This can be easily calculated by solving the following constrained problem:
\begin{equation}
\mbox{maximise} \ \Delta\bd{z}^T \bd{F} \Delta\bd{z} \ \ \  \mbox{over} \ \Delta\bd{z}, \ \   \mbox{subject to} \  \|\Delta\bd{z}\|^2=1. 
\end{equation} 
The solution to this problem is given by setting $\Delta\bd{z}^{*}$ to the eigenvector corresponding to the largest eigenvalue $\lambda^{*}$.
Eigenvalue $\lambda^{*}$ informs us of the maximum local distortion in distance and can be taken as a measure for the local magnification factor.

\section{Numerical Experiments}
\label{sec:experiments}

In the following we compare the proposed method to other visualisation algorithms  and discuss the results.

\subsection{Datasets}

In order to judge whether a visualisation truly captures similarities, we need to know a priori 
which time series are similar to which. We therefore
 employ the following particular datasets whose data items fall under known classes and are labelled. For these datasets, there is a very strong a priori expectation that the classes {\it are governed by  qualitatively distinct dynamical regimes}.
Thus, time series of the same class are expected to appear similar (close together)
in the visualisation, while time series belonging to different classes are expected to appear dissimilar (separate) in the visualisation.

\paragraph{\bf NARMA} 
We generate $100$ sequences of length $1000$ from the three qualitatively different NARMA classes \cite{Rodan2011} of orders $10, 20, 30$. The NARMA time series is an interesting benchmark problem  due to the presence of long-term dependencies.

\paragraph{\bf Cauchy}
We sample sequences from a stationary Gaussian process
with correlation function given by $c(x_t,x_{t+h})=(1+ |h|^a)^{-\frac{a}{b}}$  \cite{Gneiting2004}.
We generated $4$ classes by permuting 
parameters $a \in \{0.65,1.95\}$ and $b \in \{0.1,0.95\}$. 
We generated from each class $100$ time series of length $1,000$.
Parameters $a$ and $b$  are respectively related
to the fractal dimension (measuring self-similarity) and the Hurst
coefficient (measuring long-memory dependence) of the series.
By construction, the four classes have  distinct characteristics.

\paragraph{\bf X-ray}

The binary system GRS1915+105 is composed of an extremely heavy stellar black hole and a low-mass star. 
Material is transferred from the star towards the black hole through the 
Roche lobe. While falling into the gravitational potential of the black hole, energy is released 
by radiating X-ray and radio (jet) emission which is typical for the class of microquasars. 
A thorough investigation
carried out in \cite{2000AA}, detected the presence of
classes of distinct dynamical patterns.
Due to the lack of multiple time sequences per state, we split the observations
into equal-length parts, resulting in $161$ sequences. Here we visualise classes $delta, kappa, phi, rho$ and $chi$.

\paragraph{\bf Wind}

We visualise wind data\footnote{Kindly provided by the 
Deutscher Wetterdienst, \href{ftp://ftp-cdc.dwd.de/}{ftp://ftp-cdc.dwd.de/ \ .}} taken from the vicinity of  Hamburg, Frankfurt and Munich.
Around each city,  we select the 10 closest stations with a completeness of more than 99\% of hourly measured wind 
speed data between 13/01/2014 - 31/12/2014 ($8,471$ measurements per station). 
Missing data are interpolated using a spline function of the $3rd$ degree. In order to increase the number
of visualised entities, the time series of each station are cut into two non-overlapping parts of $4,000$ data points each. 
In these data there is a strong a priori expectation that time series
associated with the coastal city of Hamburg are different to the other data.

\paragraph{{\bf Textual data (symbolic)}}

We visualise the first chapter of 
J. K. 
Rowling's ``Harry Potter and the Philosopher's Stone'' in three languages  German, English and 
Spanish. A full symbolic representation of the alphabet makes the optimisation of the ESN 
difficult and it would be a trivial task to separate the languages as they could be identified 
by single words. Here, we choose a binary representation where the states $0$ and $1$
represent vowels and consonants. 
Punctuation and whitespaces are ignored. E.g. 
a German sentence 
is converted as follows:
\begin{center}
$\gg$Die Potters, das stimmt, das hab ich geh\"ort -$\ll$ \\
\_011\_0100100\_\_010\_001000\_\_010\_010\_100\_010100\_\_\_
\end{center}
%

Discarded symbols are marked by an underscore. This representation returns sequences of different length 
for each language, but all with at least $24,000$ symbols. To increase the number of sequences per 
language, we split the binary vectors into sequences of length $2,000$ with neighbouring sequences overlapping by $50\%$. 
It is interesting to see whether texts originating from different languages still retain their distinguishing  dynamics after subjected to this drastic ``binarisation".

\subsection{Dimensionality Reduction Algorithms}

\begin{figure*}[!t]
\centering
\subfloat[PCA on NARMA.]{%
\includegraphics[width=0.25\textwidth]{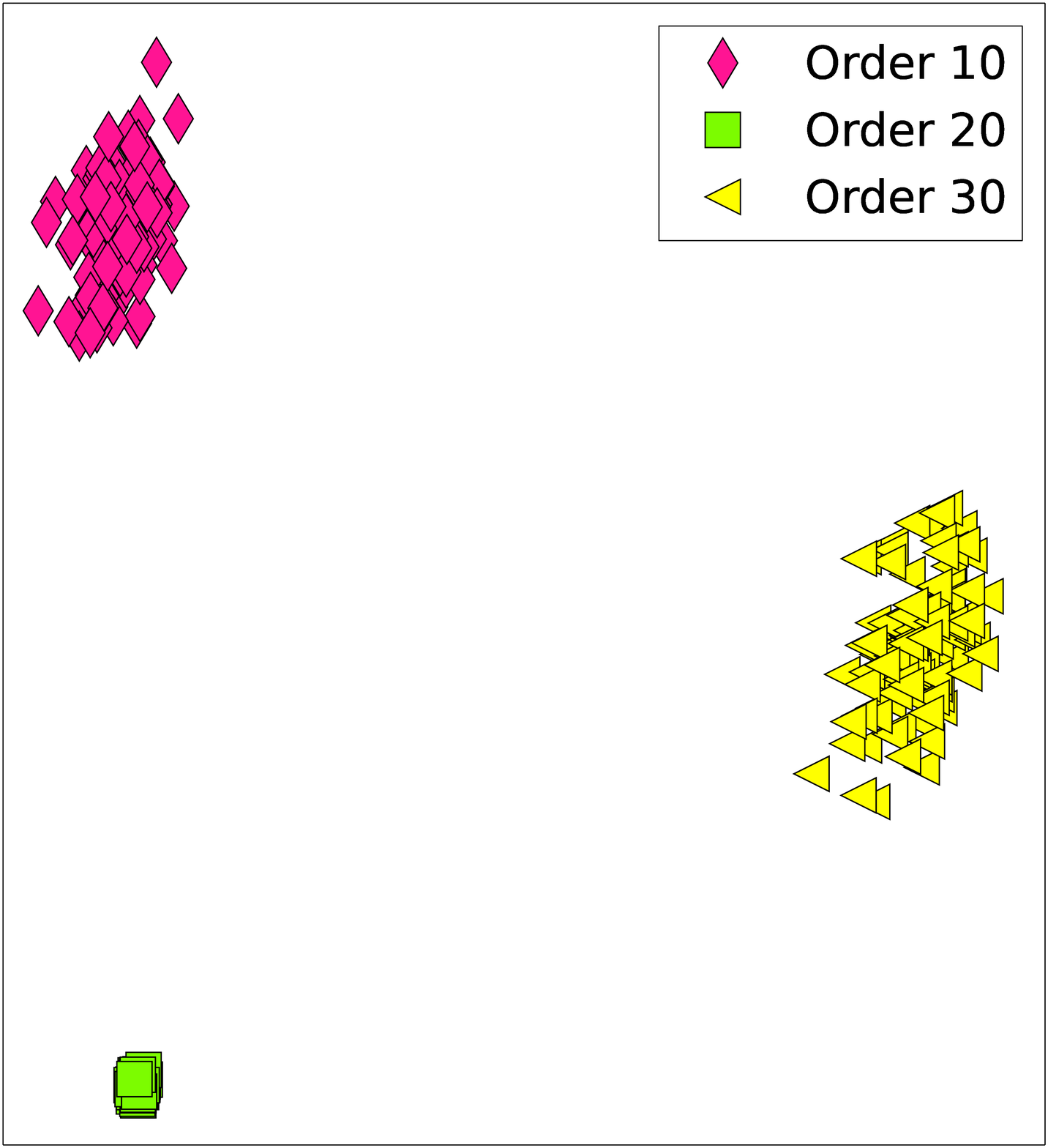}
\label{fig:pca:narma}
}
\subfloat[t-SNE on NARMA, perpl.=40.]{%
\includegraphics[width=0.25\textwidth]{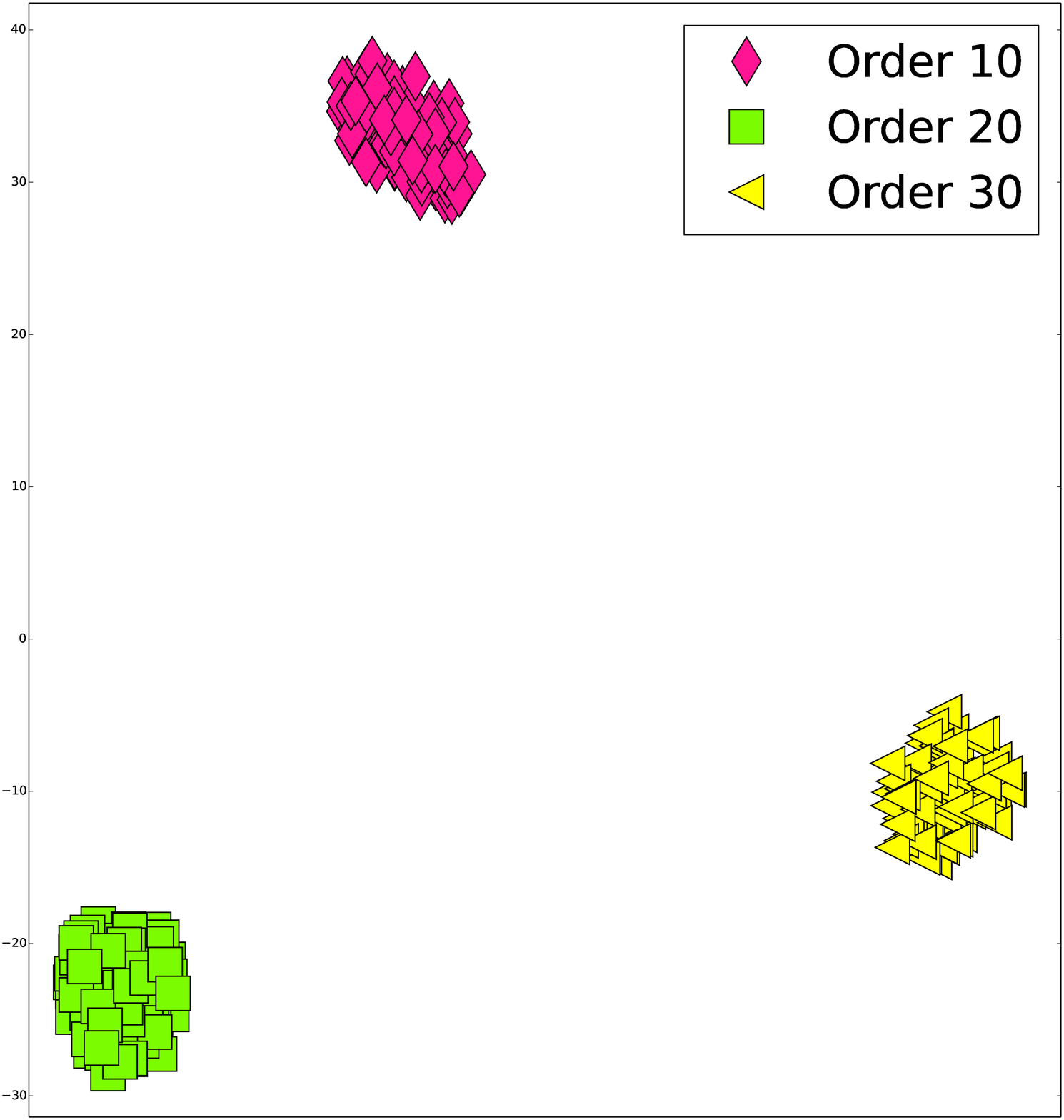}
\label{fig:tsne:narma}
}
\subfloat[Standard-AE on NARMA.]{%
\includegraphics[width=0.25\textwidth]{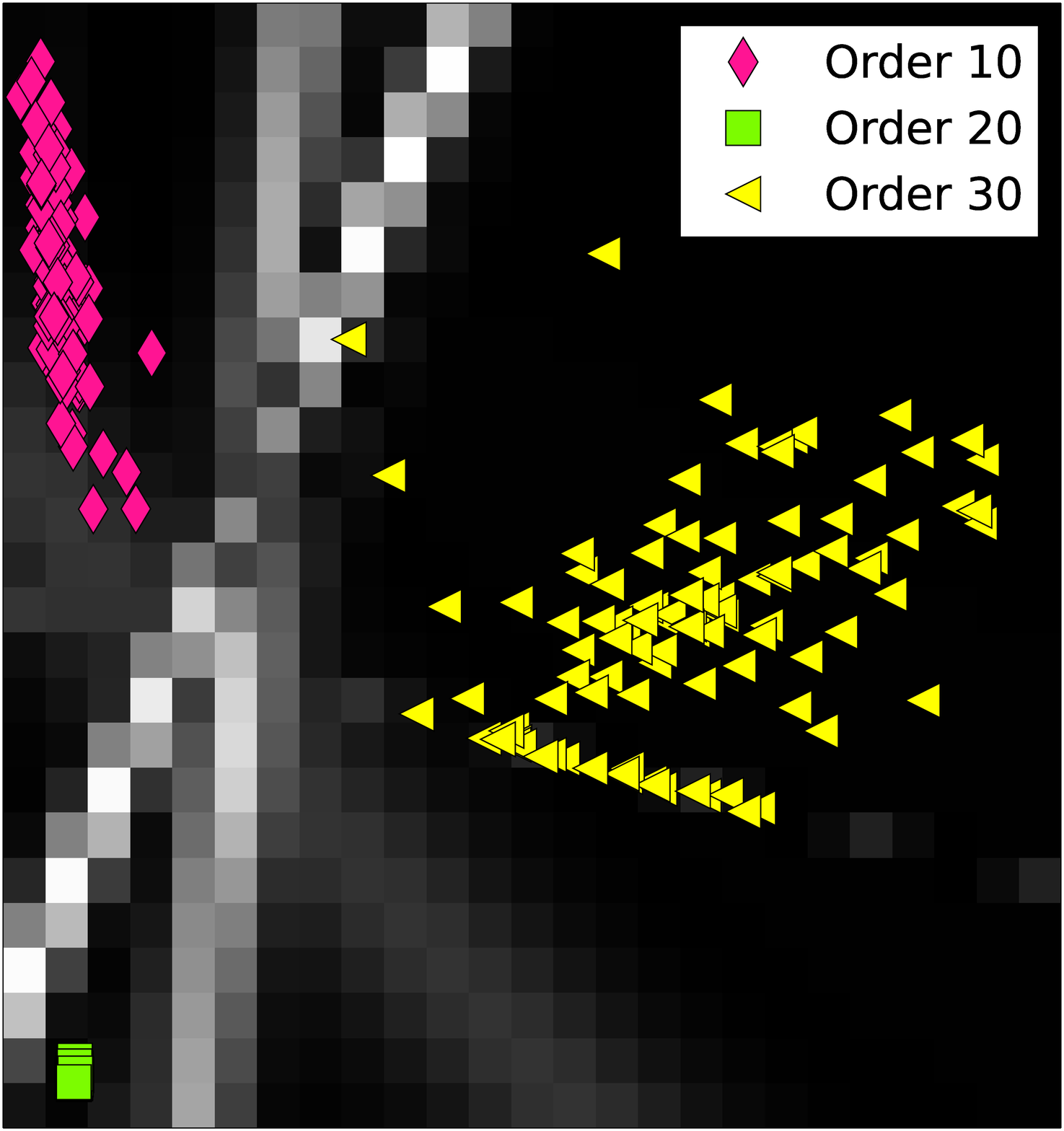}
\label{fig:ae:narma}
}
\subfloat[ESN-AE on NARMA.]{%
\includegraphics[width=0.25\textwidth]{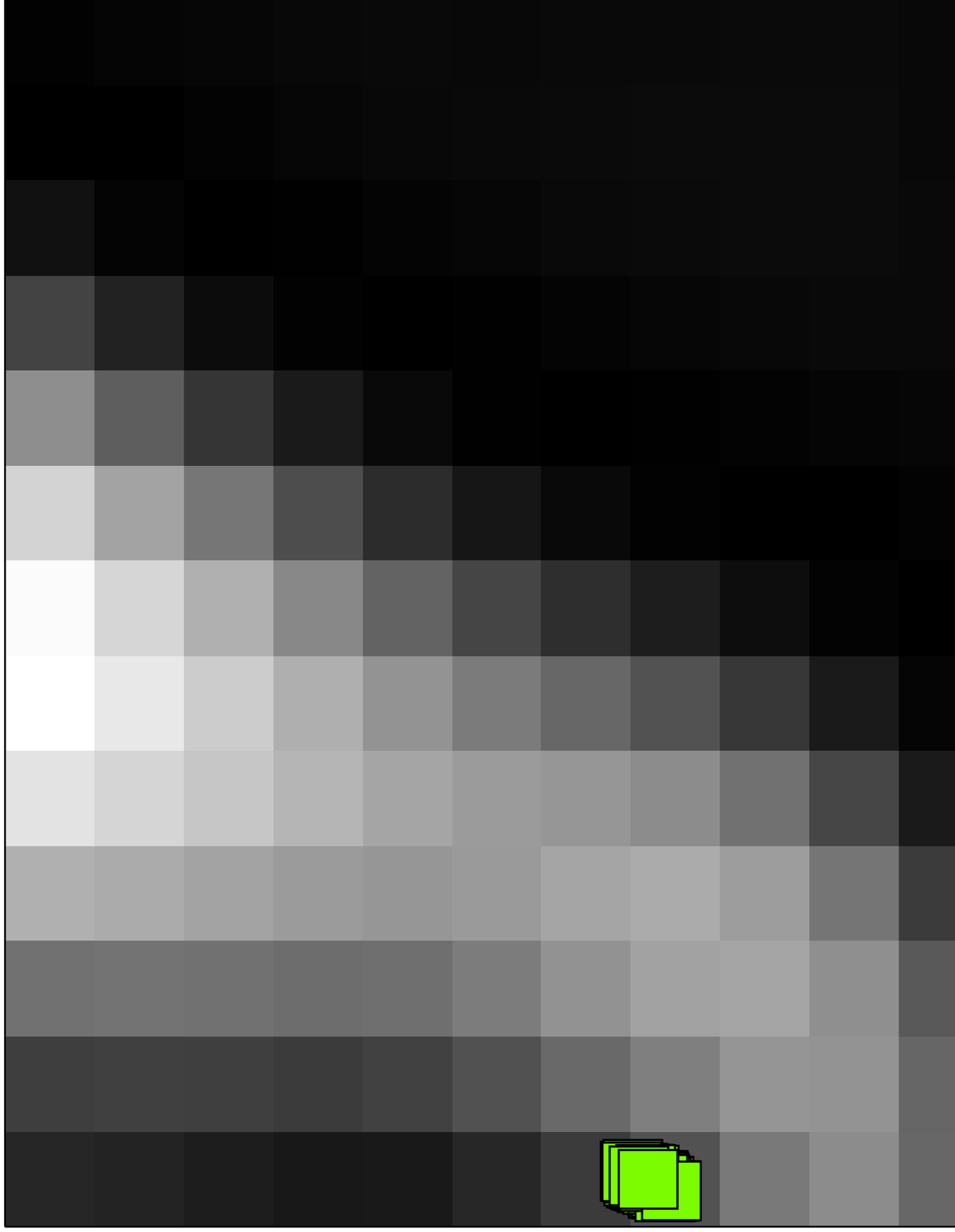}
\label{fig:esn:narma}
}
\\
\subfloat[PCA on Cauchy.]{%
\includegraphics[width=0.25\textwidth]{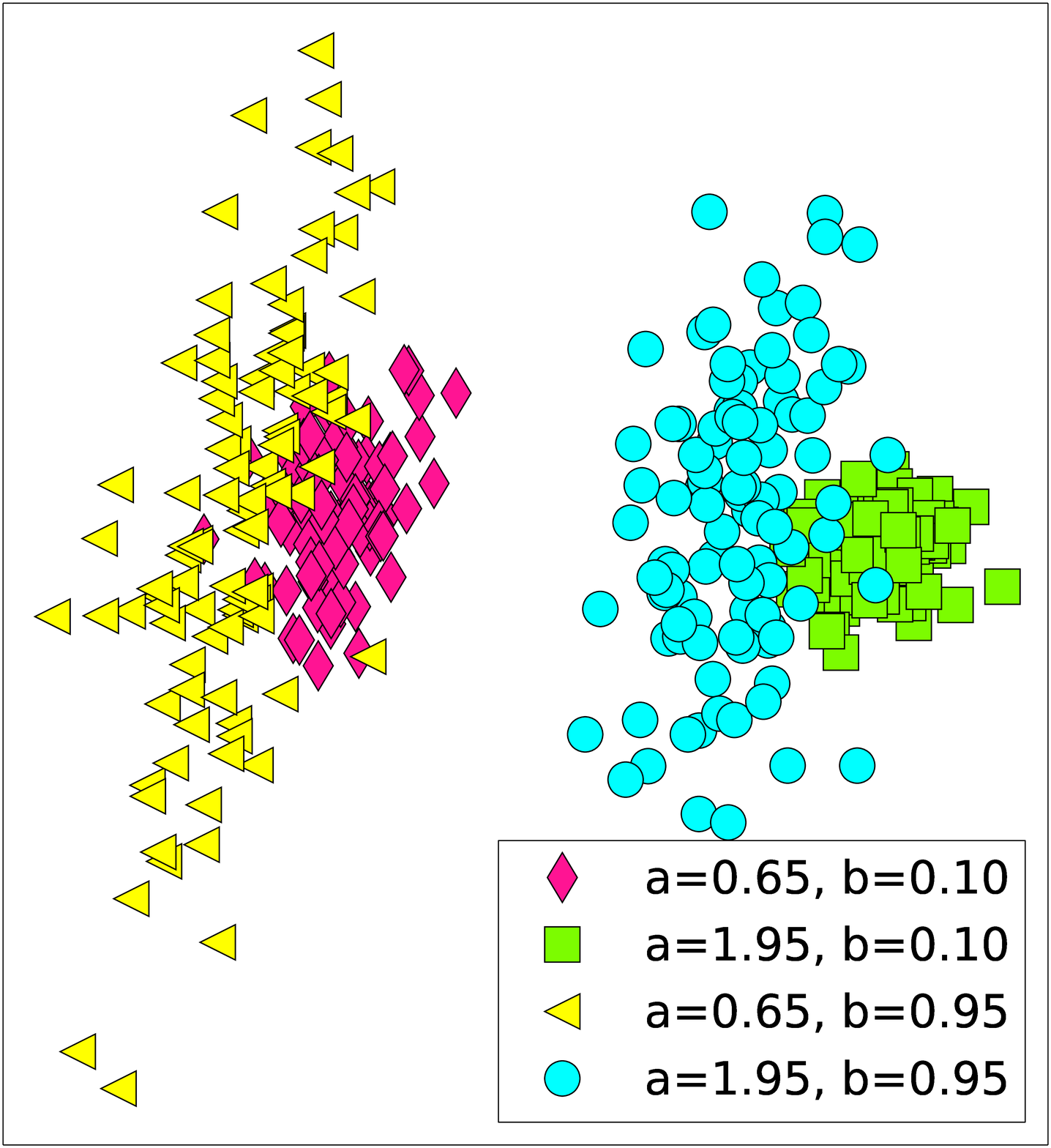}
\label{fig:pca:cauchy}
}
\subfloat[t-SNE on Cauchy, perpl.=30.]{%
\includegraphics[width=0.25\textwidth]{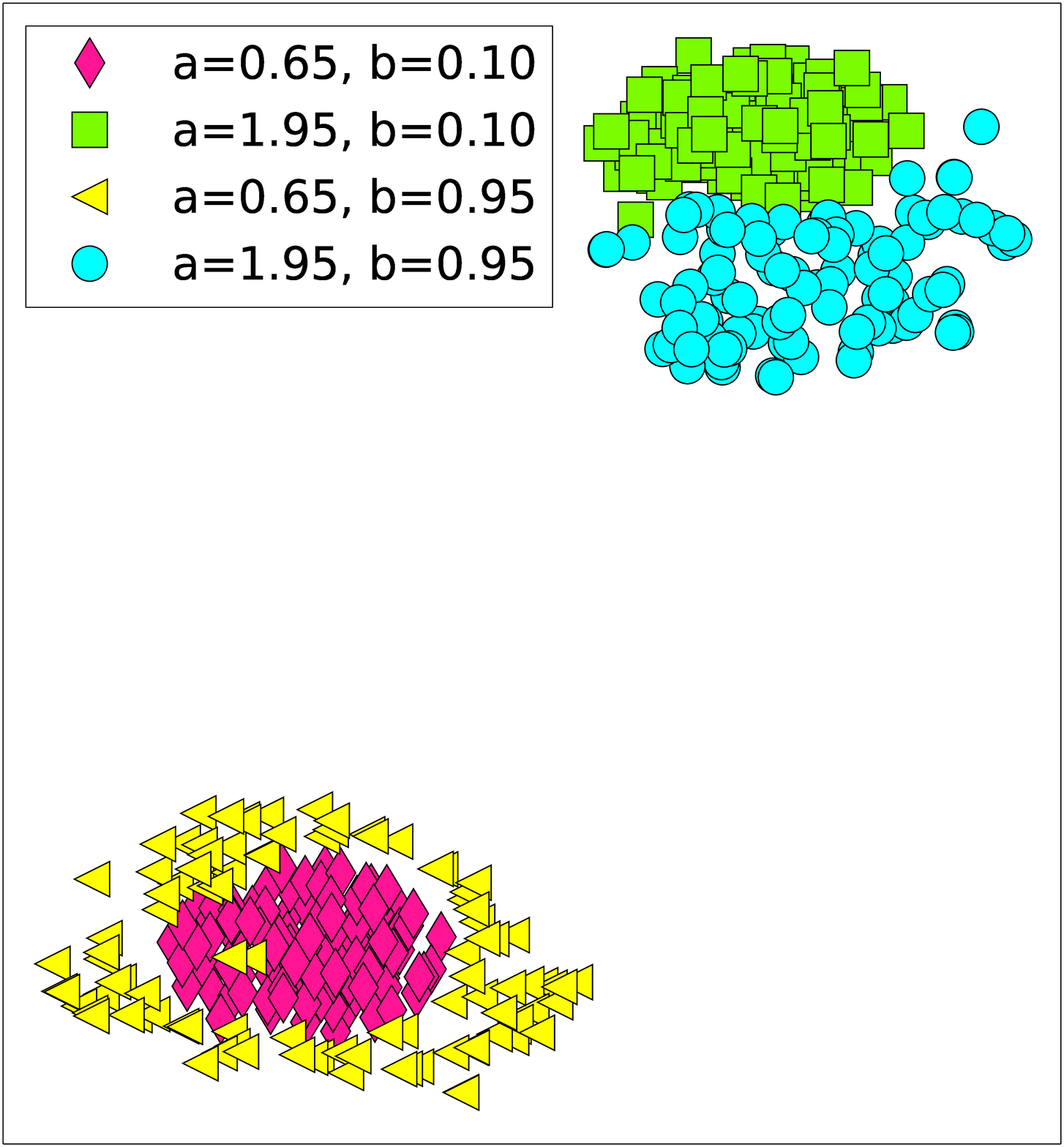}
\label{fig:tsne:cauchy}
}
\subfloat[Standard-AE on Cauchy.]{%
\includegraphics[width=0.25\textwidth]{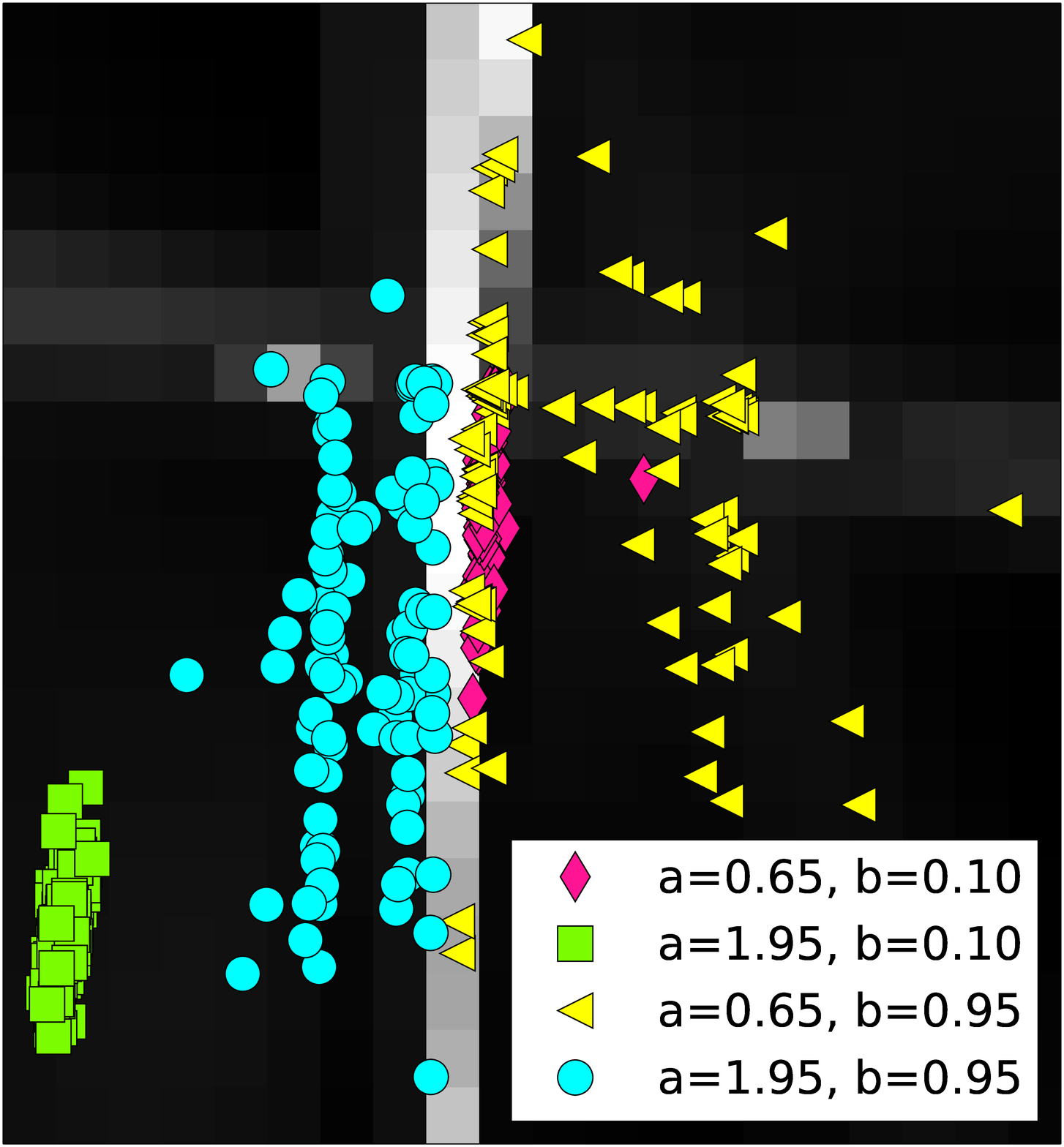}
\label{fig:ae:cauchy}
}
\subfloat[ESN-AE on Cauchy.]{%
\includegraphics[width=0.25\textwidth]{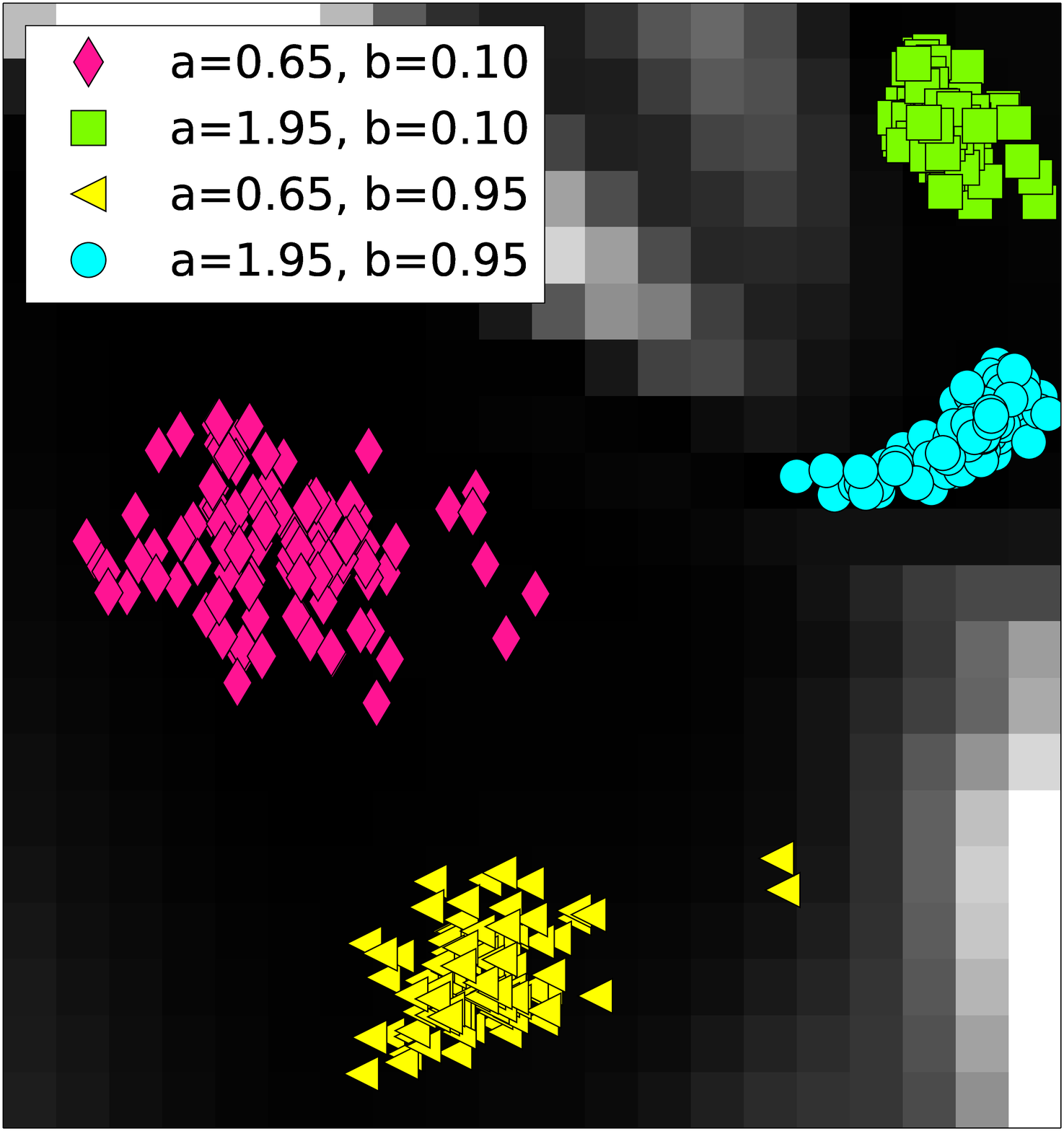}
\label{fig:esn:cauchy}
}
\\
\subfloat[PCA on X-ray.]{%
\includegraphics[width=0.25\textwidth]{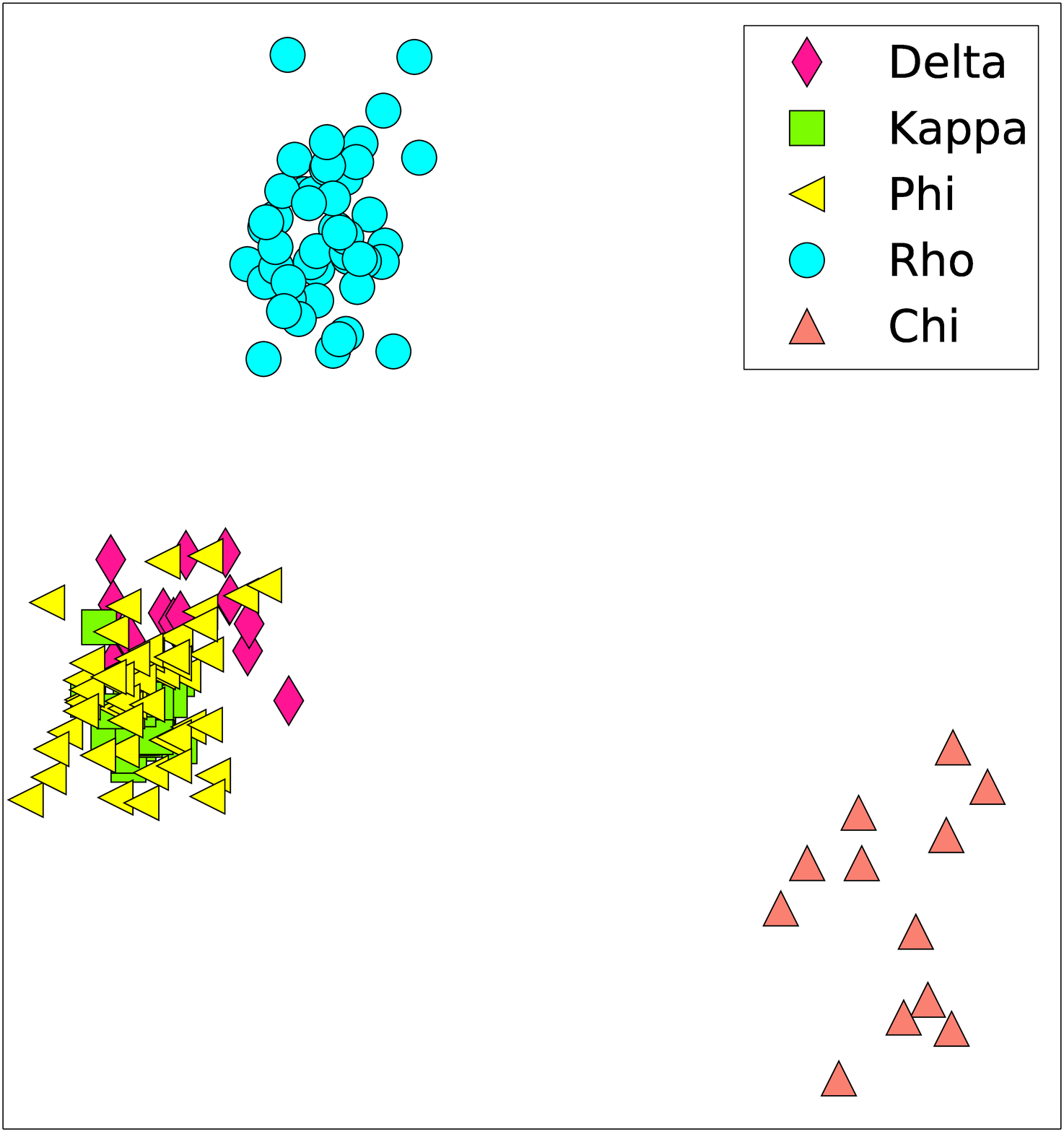}
\label{fig:pca:xray}
}
\subfloat[t-SNE on X-ray, perpl.=10.]{%
\includegraphics[width=0.25\textwidth]{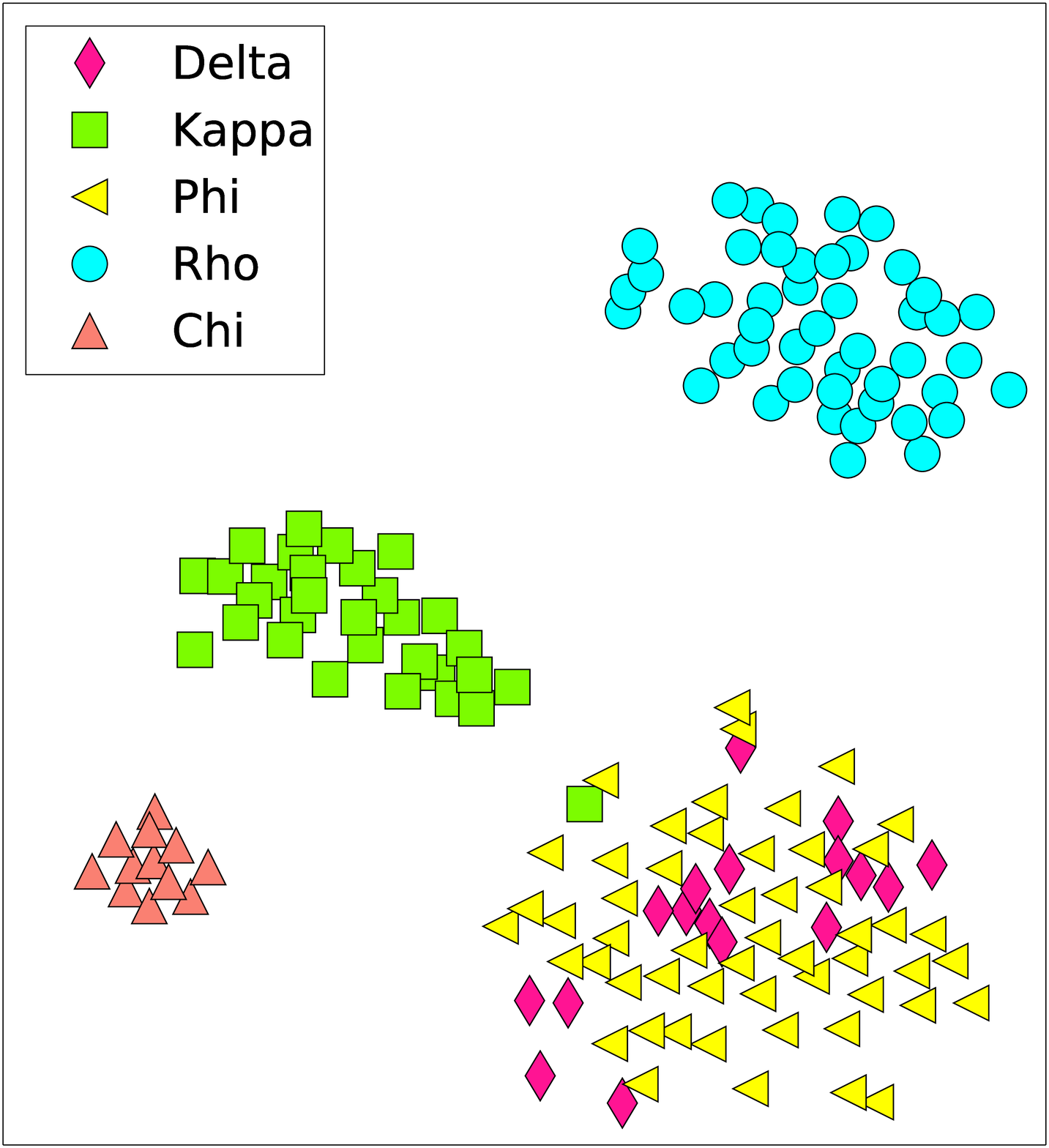}
\label{fig:tsne:xray}
}
\subfloat[Standard-AE on X-ray.]{%
\includegraphics[width=0.25\textwidth]{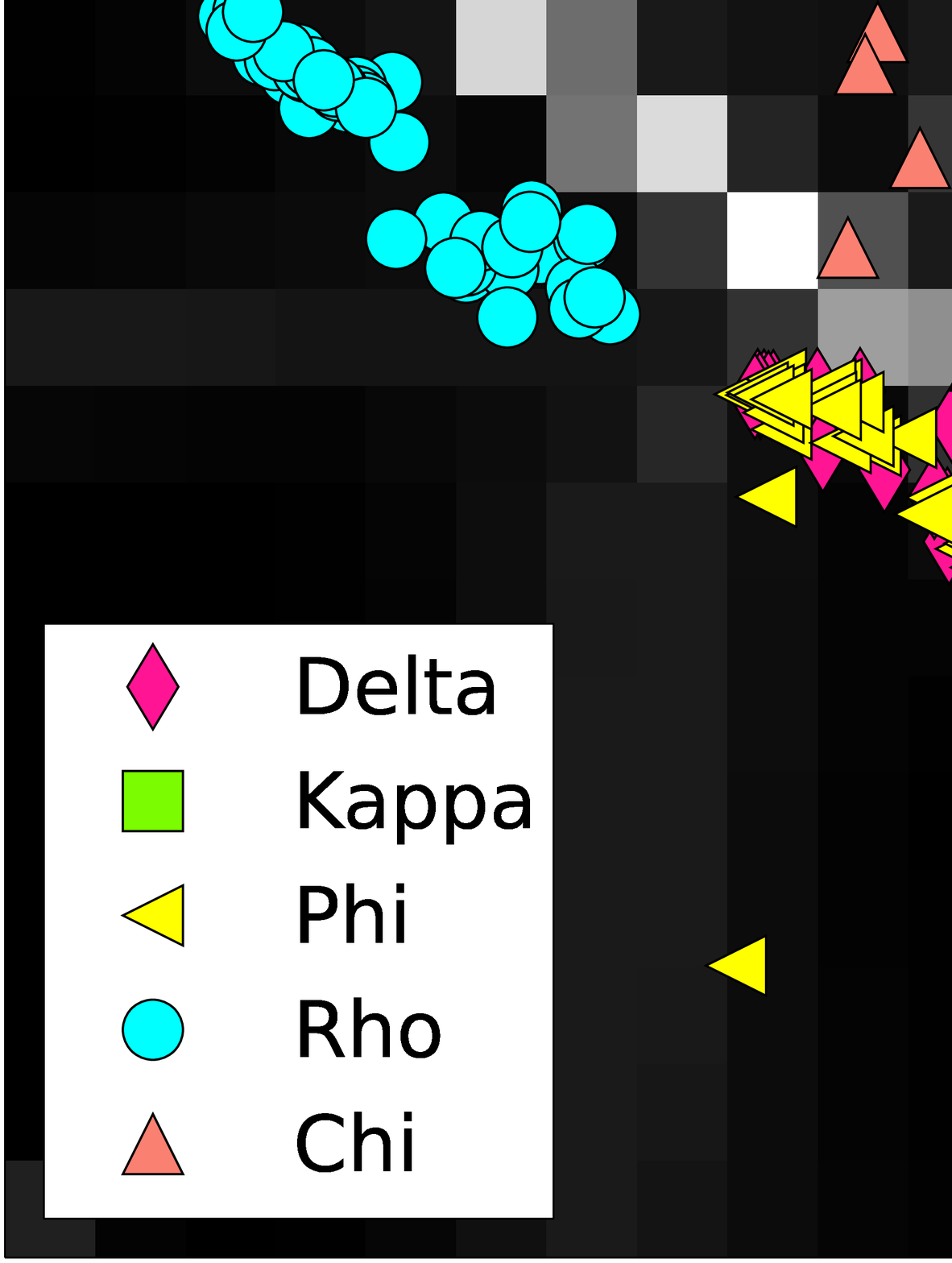}
\label{fig:ae:xray}
}
\subfloat[ESN-AE on X-ray.]{%
\includegraphics[width=0.25\textwidth]{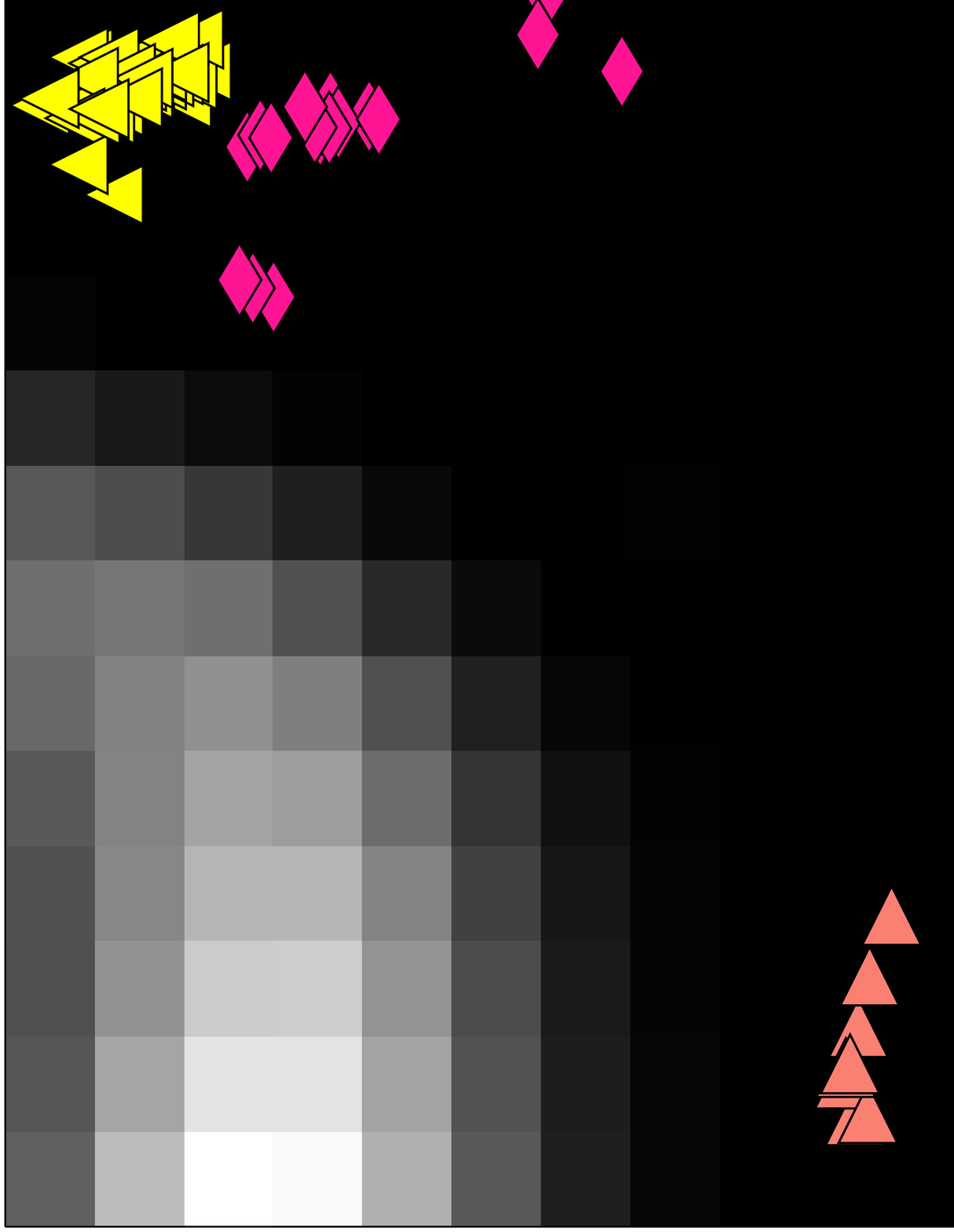}
\label{fig:esn:xray}
}
\caption{Visualisations on NARMA (top), Cauchy (middle) and X-ray (bottom) data. High/low magnifications correspond to bright/dark regions. Legends specify which markers correspond to which classes.}
\label{fig:visualisations_A}
\end{figure*}

The following dimensionality reduction algorithms are compared in the numerical experiments. 
All  algorithms operate on the readout weights $\bd{w}$. Sequences are represented as readout weights using a deterministic cyclic ESN whose
parameters are selected using the validation procedure  in Section \ref{sec:encoding}.
Additionally, in this validation scheme we include the regularisation parameter $\mu \in \{10^{-2},10^{-3},10^{-4}\}$.
In all experiments the size of the reservoir is fixed to $D=50$ and we set a washout period of $50$ time steps. We set $Q=2$ for constructing  2D visualisations.

\paragraph{\bf PCA} 
We include PCA as it helps us gauge how difficult it is to project a dataset to low dimensions: if PCA delivers a good result, this hints that a complex, non-linear projection is superfluous.

\paragraph{\bf t-SNE} 
We include t-SNE \cite{Maaten2008} as one of the most popular and well performing algorithms designed for vectorial data. We train t-SNE with perplexities in $[5,10,20,30,40,50]$, and display the visualisation that shows the best class separation. The chosen perplexity is quoted in the figures.

\paragraph{\bf Standard autoencoder (standard-AE)} 
We employ the standard autoencoder  operating directly on the readout weights.
The hidden layers of the encoding and decoding part have the same number $H$ of neurons. We also add a regulariser on the weights of the autoencoder $\nu^2 \|\bd{\theta}\|^2$ to control complexity.  In all experiments, we set $H=10$,  $\nu=1$.

\paragraph{\bf Proposed approach (ESN-AE)} 
The proposed ESN-AE has the same hyperparameters as the standard-AE. We again  fix 
the hyperparameters to $H=10$,  $\nu=1$.

\subsection{Results}

\begin{figure*}[!t]
\centering

\subfloat[PCA on Wind.]{%
\includegraphics[width=0.25\textwidth]{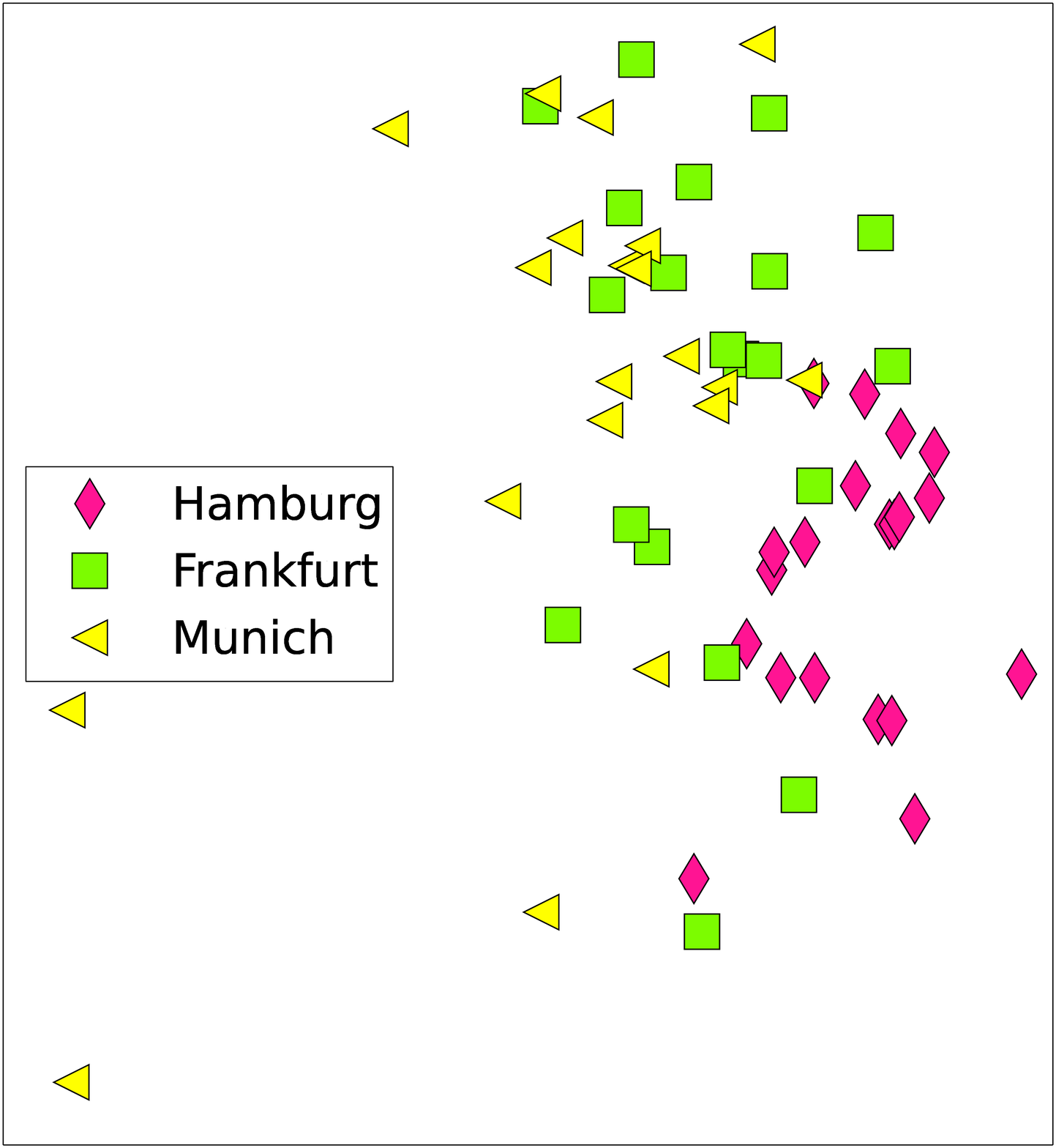}
\label{fig:pca:wind}
}
\subfloat[t-SNE on Wind, perpl.=5.]{%
\includegraphics[width=0.25\textwidth]{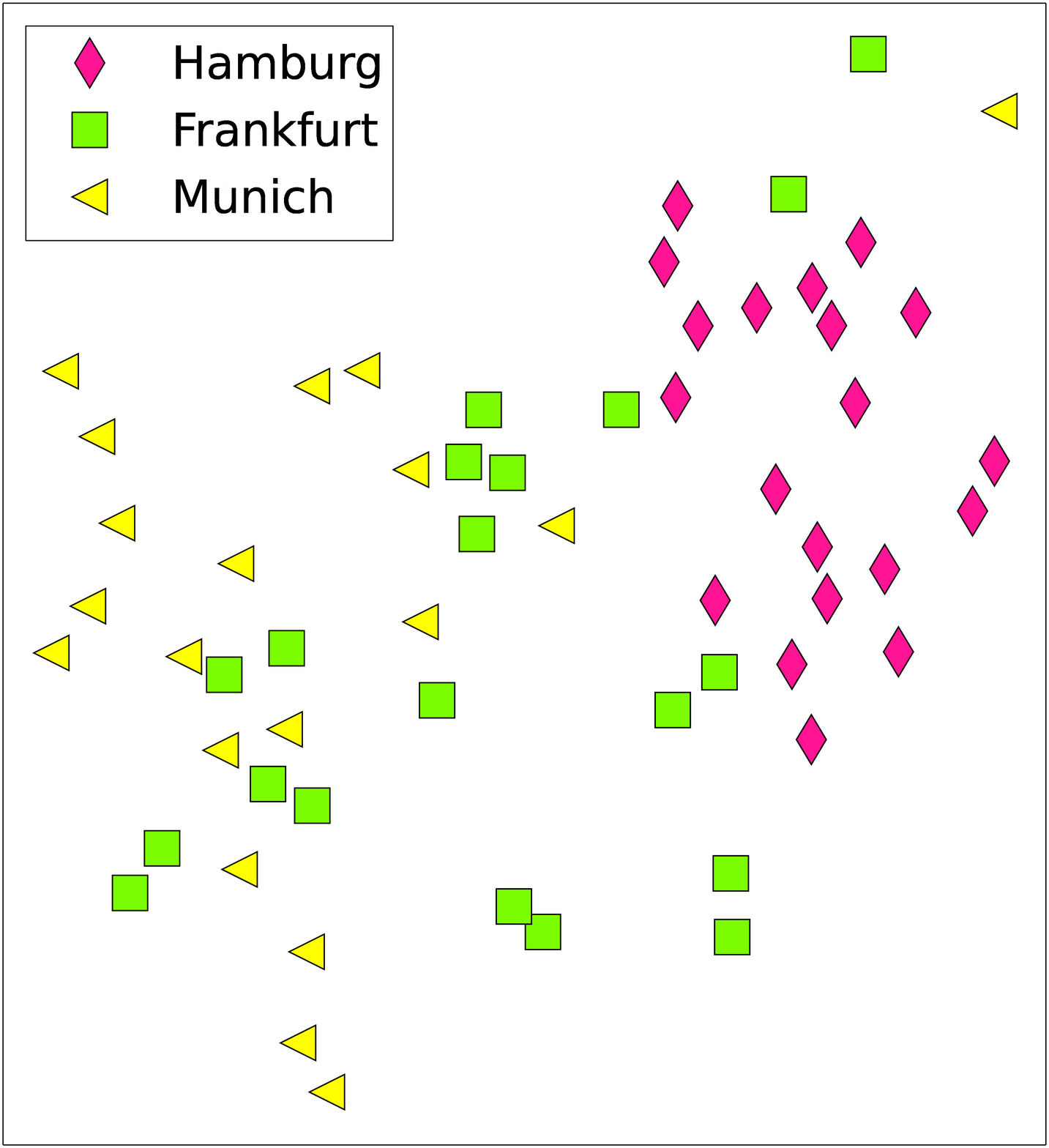}
\label{fig:tsne:wind}
}
\subfloat[Standard-AE on Wind.]{%
\includegraphics[width=0.25\textwidth]{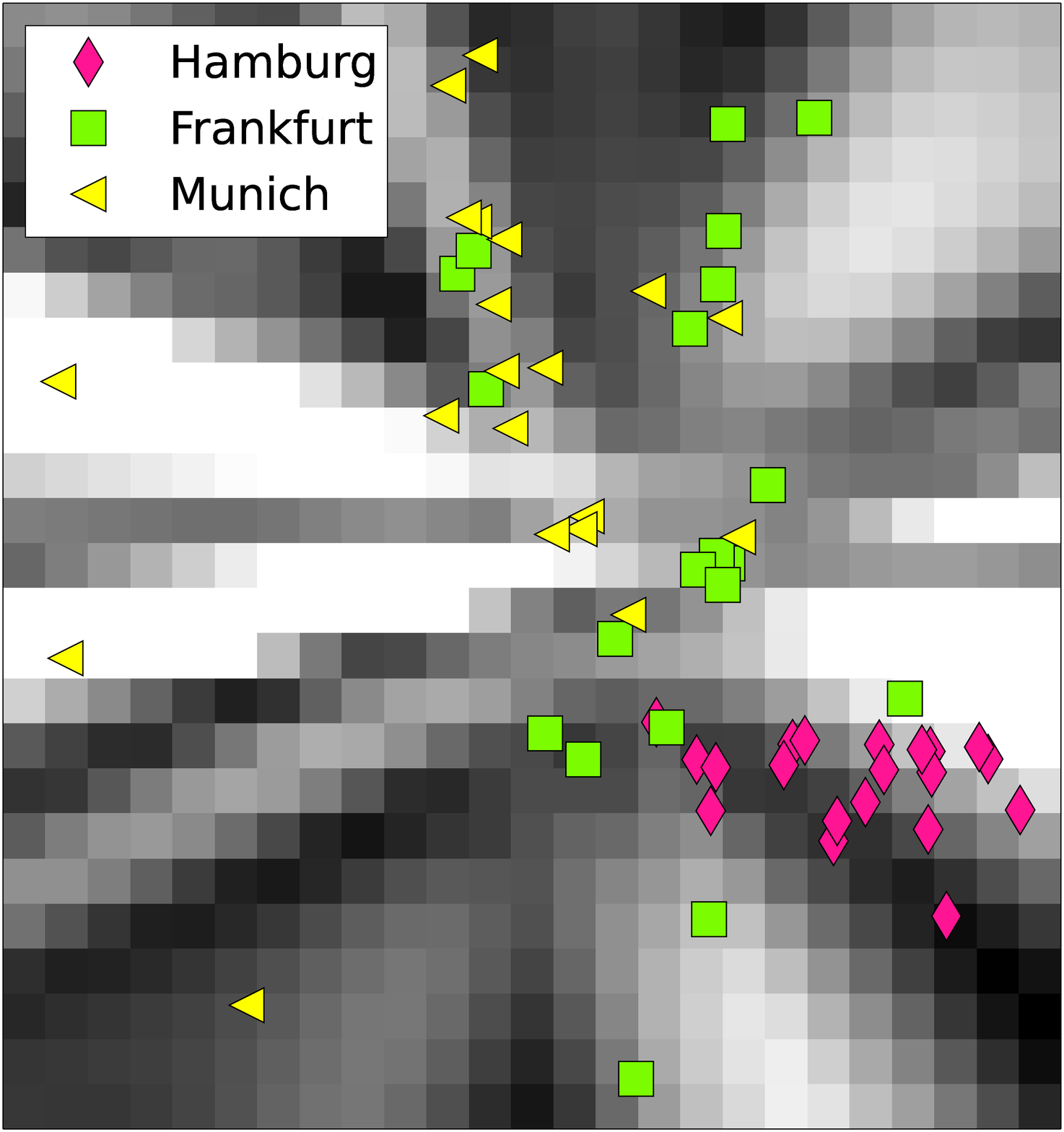}
\label{fig:ae:wind}
}
\subfloat[ESN-AE on Wind.]{%
\includegraphics[width=0.25\textwidth]{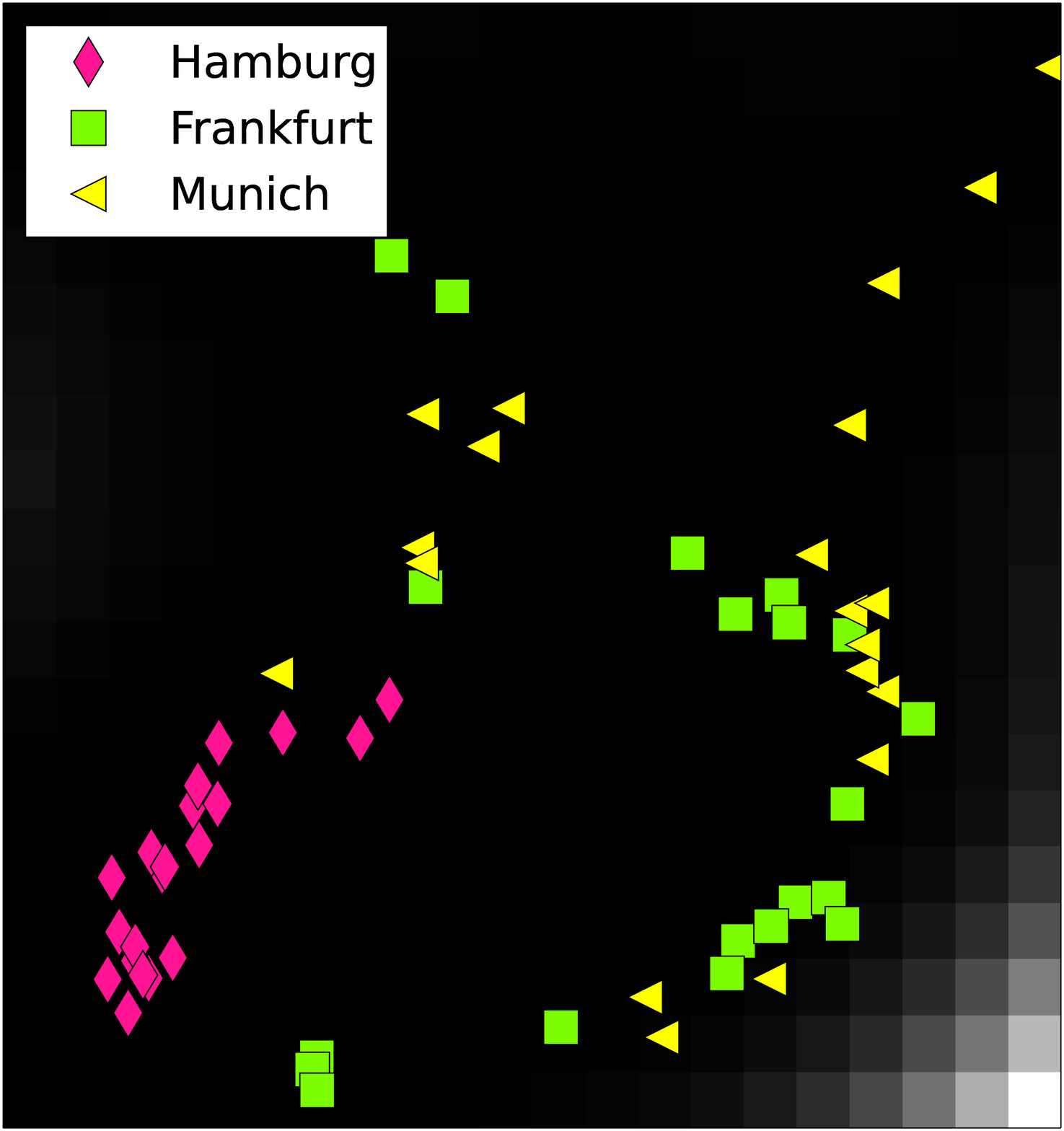}
\label{fig:esn:wind}
}
\\
\subfloat[PCA on Textual.]{%
\includegraphics[width=0.25\textwidth]{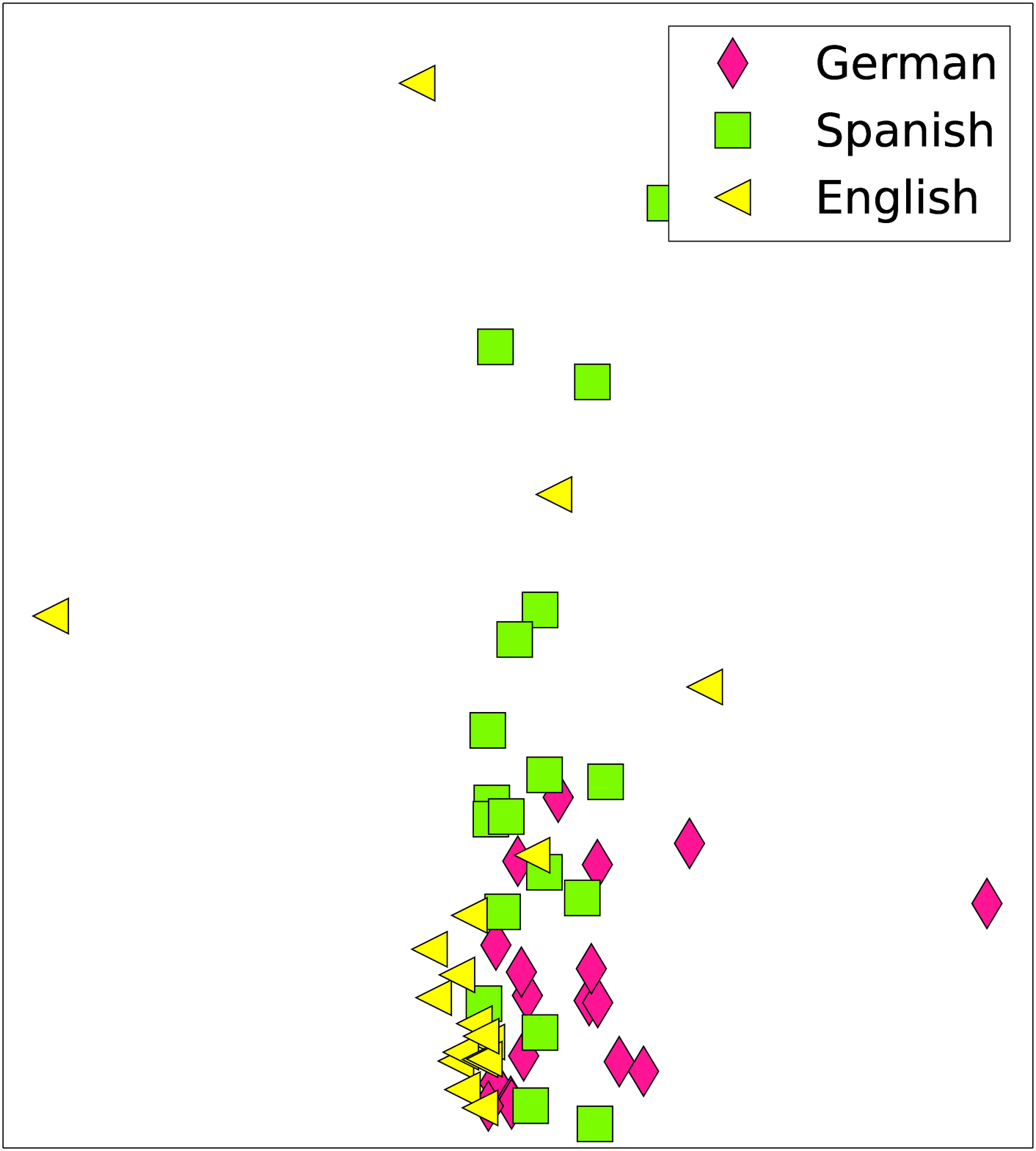}
\label{fig:pca:text}
}
\subfloat[t-SNE on Textual, perpl.=10.]{%
\includegraphics[width=0.25\textwidth]{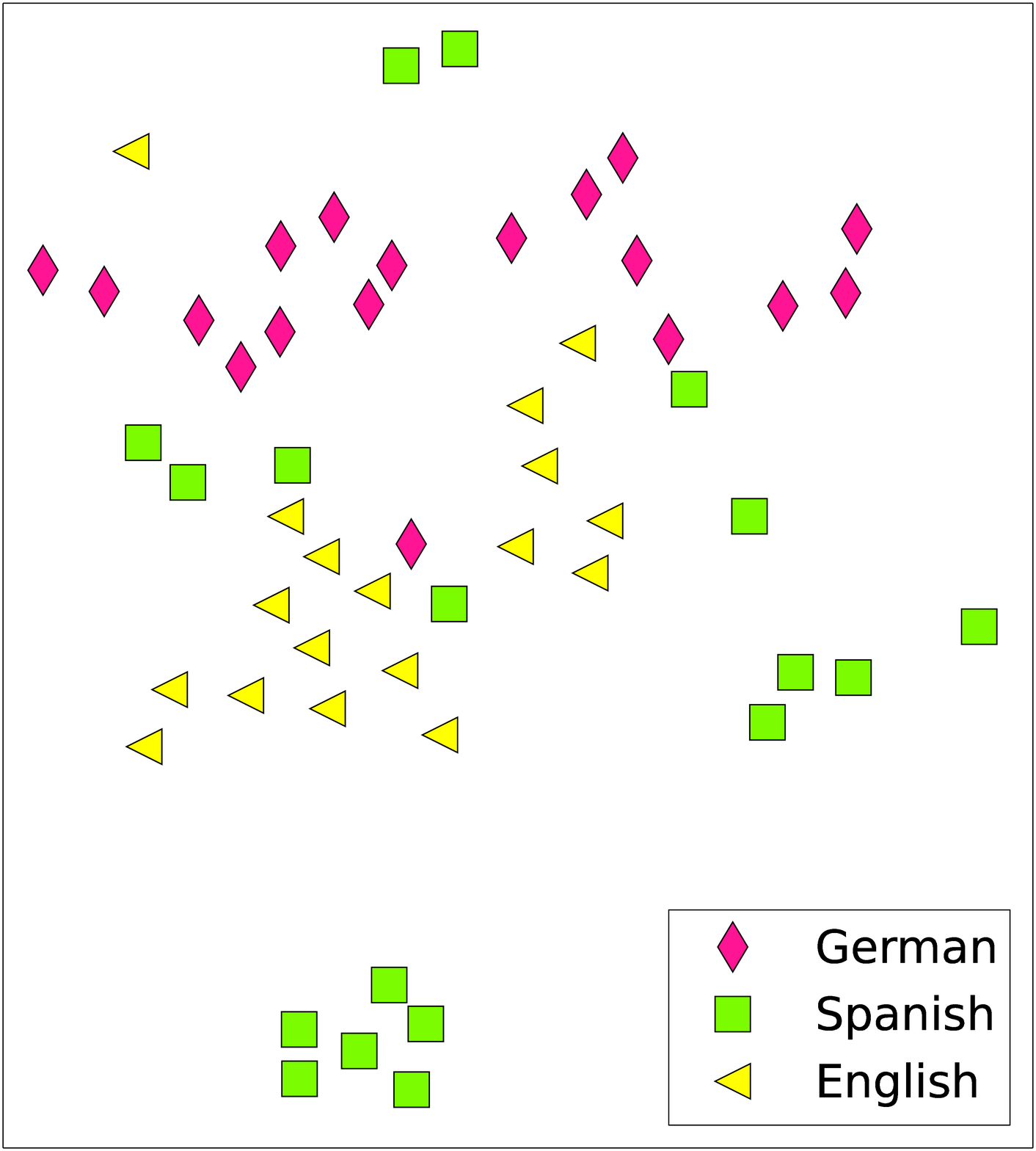}
\label{fig:tsne:text}
}
\subfloat[Standard-AE on Textual.]{%
\includegraphics[width=0.25\textwidth]{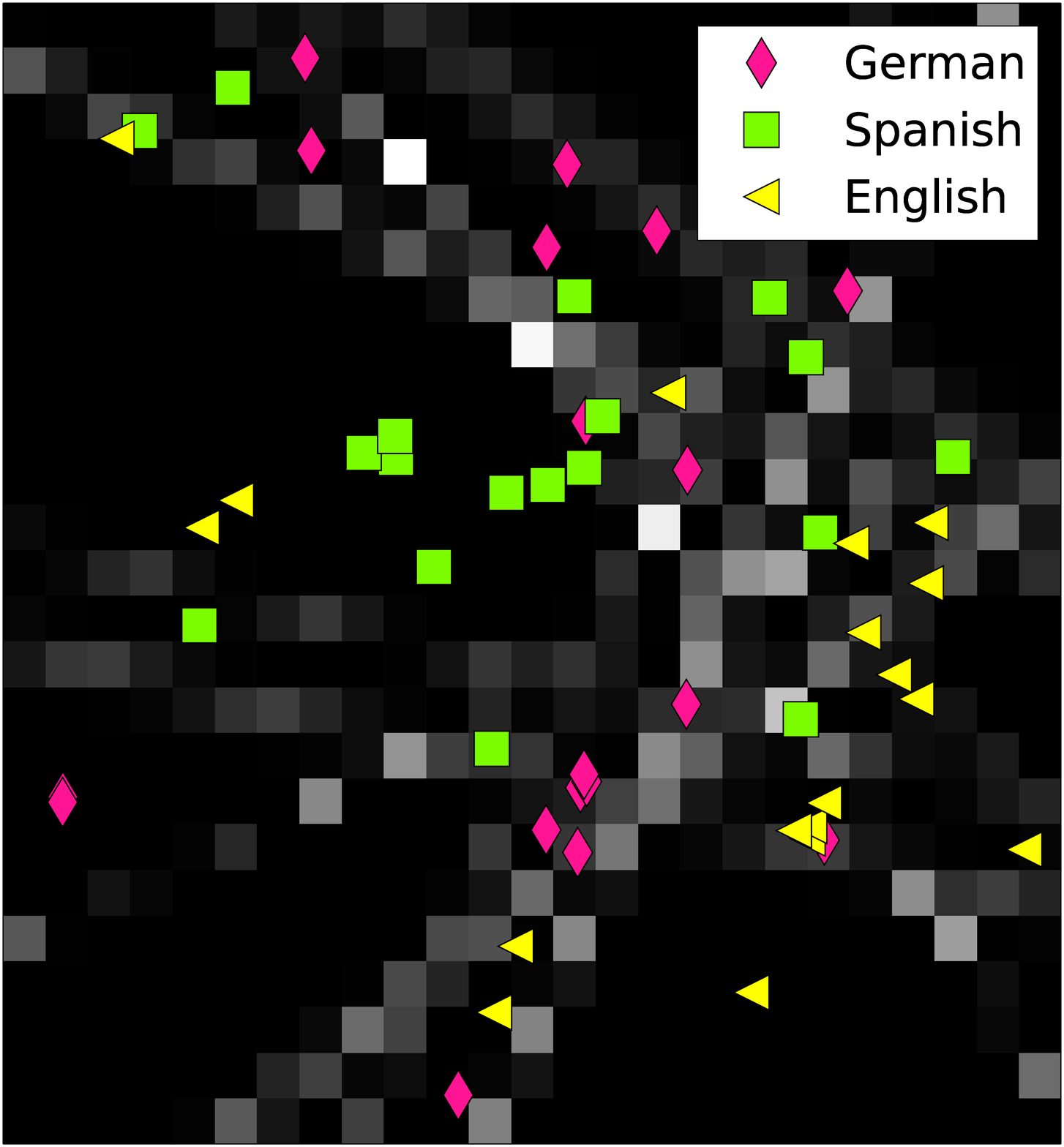}
\label{fig:ae:text}
}
\subfloat[ESN-AE on Textual.]{%
\includegraphics[width=0.25\textwidth]{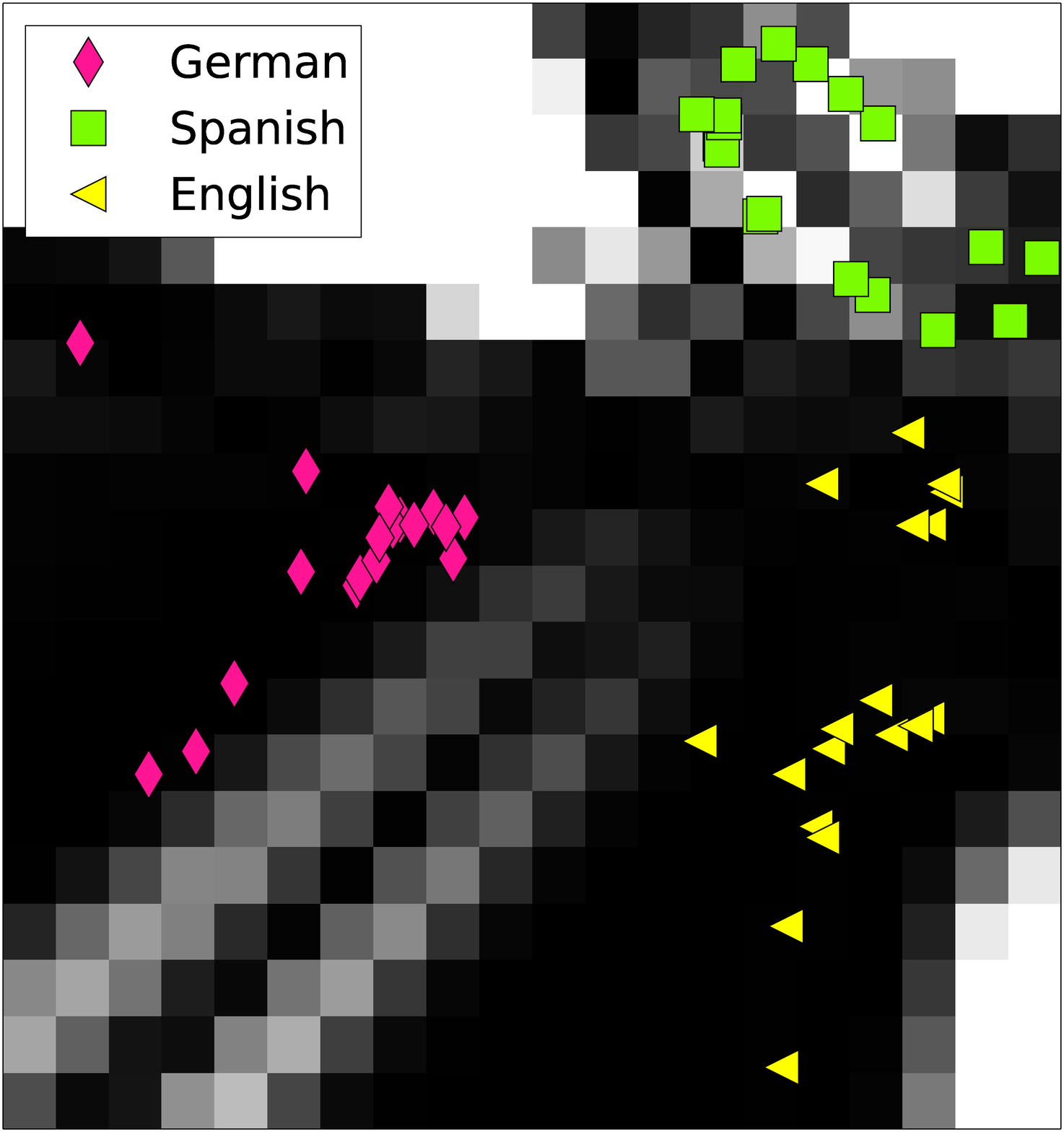}
\label{fig:esn:text}
}
\caption{Visualisations on Wind (top), and Textual (bottom) data. High/low magnifications correspond to bright/dark regions. Legends specify which markers correspond to which classes.}
\label{fig:visualisations_B}
\end{figure*}

We present the visualisations in Figs.~\ref{fig:visualisations_A} and \ref{fig:visualisations_B}. Each column of plots corresponds to one
of the aforementioned dimensionality reduction algorithms,
and each row to a dataset. The projections in the plots
appear as coloured markers of different shapes indicating class origin.
The legend in each plot shows which marker corresponds to which class.
Following Section \ref{sec:magnification}, we display local magnification factors, for the autoencoders,
as the maximum eigenvalue $\lambda^{*}$ of matrix $\bd{F}(\bd{z})$ on a regular grid of points
\bd{z} on the visualisation space.
Dark and bright values signify low and high eigenvalues/magnification factors respectively.
There are no magnification factors for PCA, as the linear mapping connecting the visualisation space
to the high-dimensional space is  length/distance preserving. Also, we do not present magnification factors for t-SNE, as it does not define an explicit mapping between the visualisation and high-dimensional space. It thus requires a different framework than the one used here in order to study magnifications.

\begin{table}[]
\centering
\caption{Mean squared errors between NARMA classes, the smaller the more similar. }
\label{tbl:narma_cross_loglikel}
\begin{tabular}{lrrr}
\hline
         & \multicolumn{1}{l}{\small Order 10} & \multicolumn{1}{l}{\small Order 20} & \multicolumn{1}{l}{\small  Order 30} \\ \hline
\small Order 10 &\small  5.331                        &\small  2185.935                     &\small  37.161                       \\
\small Order 20 &\small  2213.019                     &\small  0.052                        & \small 2030.409                     \\
\small Order 30 &\small  30.478                       &\small  1983.585                     &\small  6.031                        \\ \hline
\end{tabular}
\end{table}

\paragraph{{\bf NARMA}, top row in Fig.~\ref{fig:visualisations_A}}

We note  that all visualisations separate the three classes, and  show that the three classes are equidistant. The magnifications in \ref{fig:ae:narma} show that the standard-AE views the three classes indeed as distinctly separable clusters. However, in the case of the ESN-AE in \ref{fig:esn:narma}, the magnification factors suggest the presence of  distortions in  distances close to class ``Order 20''. This means that in actual fact  class ``Order 20''  is separated by significant distance from the other two classes, and that classes ``Order 10'' and ``Order 30'' are closer and more similar to each other.  We investigate this hypothesis with a simple experiment. We generate from each class additional $200$ sequences. For each pair of classes (classes also pair with themselves), we train on sequences from one class and measure the mean squared error on the unseen sequences of the other classes. These errors are reported in Table \ref{tbl:narma_cross_loglikel}, and support this hypothesis put forward by the magnification factors in the ESN-AE visualisation.

\paragraph{{\bf Cauchy}, middle row in Fig.~\ref{fig:visualisations_A}}
PCA in \ref{fig:pca:cauchy} and t-SNE in \ref{fig:tsne:cauchy} manage to organise the classes coherently to some degree, while the standard-AE in \ref{fig:ae:cauchy}
fails to produce a convincing result. ESN-AE in \ref{fig:esn:cauchy} displays a clear separation between all four classes.
In particular the presence of magnification factors close to the two classes located in the upper right corner, shows that these two classes are potentially separated by a larger distance to the other two.

\paragraph{{\bf X-ray}, bottom row in Fig.~\ref{fig:visualisations_A}}

All visualisations clearly separate the $rho$ and $chi$ classes.
For standard-AE in \ref{fig:ae:xray}, the strong magnification suggest that the $chi$ class is quite different to the others. t-SNE in \ref{fig:tsne:xray} and ESN-AE distinguish in \ref{fig:esn:xray} the classes in a clearer fashion. ESN-AE exhibits less overlapping projections, but does not put enough distance between  classes $delta$ and $phi$.   The presence of magnifications close to the $chi$ class is a hint that this class is quite different to the other ones.  Still even in the absence of labels (i.e. colour markers), the classes are identifiable in the visualisation produced by ESN-AE.

\begin{table}[!t]
\centering
\caption{Mean reconstruction and standard deviation, averaged over 10 runs.}
\label{tbl:reconstruction}
\begin{tabular}{lrrr}
\hline

& \multicolumn{1}{c}{\scriptsize \hspace{-0.25cm} PCA}             &  \multicolumn{1}{c}{\hspace{-0.3cm} \scriptsize standard AE} & \multicolumn{1}{c}{\hspace{-0.2cm}  \scriptsize ESN-AE} \\ \hline
\scriptsize \hspace{-0.25cm}NARMA              & \hspace{-0.35cm} \scriptsize 151451.130 $\pm$ 14984.801 &\scriptsize 116.041 $\pm$ 46.606     &\hspace{-0.1cm}  \scriptsize 44.126 $\pm$ 11.962  \\ \hline
\scriptsize \hspace{-0.25cm} Cauchy             & \hspace{-0.35cm} \scriptsize 121.110 $\pm$ 4.022                             &\hspace{-0.2cm}  \scriptsize 102.115 $\pm$ 2.522                     & \scriptsize 95.176 $\pm$ 2.675                            \\ \hline
\scriptsize \hspace{-0.25cm} Xray               & \hspace{-0.35cm} \scriptsize 25.376 $\pm$ 2.274                              &\hspace{-0.2cm}  \scriptsize 42.727 $\pm$ 17.423                   &\scriptsize  21.798 $\pm$ 1.617                            \\ \hline
\scriptsize \hspace{-0.25cm} Wind              & \hspace{-0.35cm} \scriptsize 5.0498 $\pm$ 0.253                              &\hspace{-0.2cm}  \scriptsize 5.192 $\pm$ 0.260                         & \scriptsize 5.079 $\pm$ 0.231                            \\ \hline
\scriptsize \hspace{-0.25cm} Textual &\hspace{-0.35cm} \scriptsize  0.691 $\pm$ 0.007   &\scriptsize 0.700 $\pm$ 0.010                         &\hspace{-0.2cm}  \scriptsize 0.694 $\pm$ 0.017                             \\ \hline

\end{tabular}
\end{table}

\paragraph{{\bf Wind}, top row in Fig.~\ref{fig:visualisations_B}}

None of the visualisations separates the Munich from the Frankfurt stations.
Matching our  prior expectation, 
ESN-AE in \ref{fig:esn:wind} organises the stations around Hamburg in a single region,
in contrast to the  other visualisations which show overlap.
Standard-AE fails to produce a clear result and its magnifications do not
help in its interpretation any further.

\paragraph{{\bf Textual data (symbolic)}, bottom row in Fig.~\ref{fig:visualisations_B}}

The binary representation of the text data in three different languages shows the true power behind the 
ESN-AE equipped here with the logistic function. While other visualisations do not exhibit  adequate separation,  ESN-AE in \ref{fig:esn:text} exhibits some clear organisation.   Additionally, magnifications suggest some separation between the German and English sequences. The bright magnifications that appear in the unpopulated corners are simply artefacts as the model has not seen any data in these areas.

\paragraph{{\bf Reconstruction}}

In order to give a quantitative impression of the quality of the visualisations,
we report reconstruction errors in Table \ref{tbl:reconstruction}. Each dataset is randomly
partitioned $10$ times into  equally sized training and test sets.
For each partitioning, we train the dimensionality algorithms and measure the
error on the test data using Eq. \eqref{eq:ESN_objective}. 
For the binary textual data, the error is measured as the fraction of predictions coinciding
with the binary test sequences. We exclude t-SNE as it does not offer a way of reconstructing
weights from the projections.

\section{Discussion}
\label{sec:discussion}

Though the conversion of time-series into fixed-length representations is not new (e.g. \cite{Grabocka2015}), we 
believe that  converting the time series via a non-parametric state space
model with fixed dynamic part (i.e~ESN) in conjunction 
with  an appropriately defined
reconstruction function, does provide a new way of performing dimensionality reduction on time series.
The results show that the proposed visualisation is better at understanding
what makes sequences (dis)similar as it 
manages to separate classes that are governed by qualitatively distinct dynamical regimes. Indeed, the produced visualisations
reflect our prior expectations as to which sequences should be similar.

Of course, combining the ESN with the autoencoder is just one possible scheme, and certainly other dimensionality reduction
schemes can be devised along this line.
One can exchange the ESN with other models such as
autoregressive-moving-average models (ARMA), and use them to cast the time-series to fixed parameter vectors.
E.g.~for slow changing signals, models based on Fourier series
might be more suitable than the ESN.
Choosing the ESN to model the temporal features of the sequences, is indeed a subjective choice. However, this does not mean that it is a bad choice: in the relevant literature, a wealth of applications
demonstrate that ESNs are good models for 
a large variety of real-world time series.

Besides the autoencoder, other dimensionality reduction methods that rely on optimising reconstruction error (e.g. GPLVM \cite{Lawrence2005}) can be adapted to the visualisation of time-series; one has to modify their objective to measure reconstruction in the sequence space, just as the loss function of ESN-AE does.

\section{Conclusion}
\label{sec:conclusion}

We have presented a method for the visualisation of time series that couples an ESN to an autoencoder. Time series are represented as readout weights of an ESN and are  subsequently compressed to a low dimensional representation by an autoencoder. The autoencoder attempts reconstruction of the readout weights in the context of the state space pertaining to the sequences thanks to the modified loss function. 
In future research, we plan to work on irregularly sampled time series that originate from eclipsing binary stars. The ESN will be replaced by a physical model that will cast the time series to vectors of physical parameters.

\section*{Acknowledgement}
The authors from HITS gratefully acknowledge the support of the Klaus Tschira Foundation.
We thank Ranjeev Misra for providing the X-ray dataset.
We would also like to thank the anonymous reviewers for helping us improve the presented work.


  \bibliographystyle{plain} 
  \bibliography{main}


%
%
%
\end{document}